\documentclass[journal,doublecolumn]{IEEEtran}
\usepackage{amsmath,amssymb,verbatim,cite,amsopn,graphicx,citesort,url,color,mathtools,epsfig,epstopdf}

\usepackage{balance}

\newcommand{\wor}{w}
\newcommand{\opt}{o}

\newcommand{\sgs}{\sigma^2}

\newcommand{\Nsate}{N_{0e}}
\newcommand{\Nsatb}{N_{0b}}
\newcommand{\Nsati}{N_{0i}}

\newcommand{\bQ}{\mathbf{Q}}
\newcommand{\bR}{\mathbf{R}}

\newcommand{\bZ}{\mathbf{Z}}
\newcommand{\bz}{\mathbf{z}}

\newcommand{\by}{\mathbf{y}}
\newcommand{\bn}{\mathbf{n}}

\newcommand{\bH}{\mathbf{H}}
\newcommand{\bh}{\mathbf{h}}
\newcommand{\bG}{\mathbf{G}}
\newcommand{\bg}{\mathbf{g}}
\newcommand{\bx}{\mathbf{x}}

\newcommand{\bw}{\mathbf{w}}
\newcommand{\bI}{\mathbf{I}}

\newcommand{\bv}{\mathbf{v}}

\newcommand{\bK}{\mathbf{K}}
\newcommand{\bk}{\mathbf{k}}

\newcommand{\lam}{\lambda}

\newcommand{\Nbm}{N_{b,\min}}

\newtheorem{Proposition}{Proposition}
\newtheorem{Theorem}{Theorem}

\newtheorem{Remark}{Remark}

\begin{document}

\title{Achieving Secrecy without Knowing the Number of Eavesdropper Antennas}

\author{
Biao~He,~\IEEEmembership{Student~Member,~IEEE,}
Xiangyun~Zhou,~\IEEEmembership{Member,~IEEE,}
and Thushara~D.~Abhayapala~\IEEEmembership{Senior Member,~IEEE}

\thanks{This work was supported by the Australian Research Council under Discovery Project Grant DP150103905.}
\thanks{B. He, X. Zhou and T. D. Abhayapala are with the Research School of Engineering, Australian National University, Canberra, ACT 0200, Australia (e-mail: biao.he@anu.edu.au, xiangyun.zhou@anu.edu.au, thushara.abhayapala@anu.edu.au).}
}

\maketitle

\vspace{-10mm}
\begin{abstract}
The existing research on physical layer security commonly assumes the number of eavesdropper antennas to be known. Although this assumption allows one to easily compute the achievable secrecy rate, it can hardly be realized in practice. In this paper, we provide an innovative approach to study secure communication systems without knowing the number of eavesdropper antennas by introducing the concept of spatial constraint into physical layer security. Specifically, the eavesdropper is assumed to have a limited spatial region to place (possibly an infinite number of) antennas. From a practical point of view, knowing the spatial constraint of the eavesdropper is much easier than knowing the number of eavesdropper antennas. We derive the achievable secrecy rates of the spatially-constrained system with and without friendly jamming. We show that a non-zero secrecy rate is achievable with the help of a friendly jammer, even if the eavesdropper places an infinite number of antennas in its spatial region. Furthermore, we find that the achievable secrecy rate does not monotonically increase with the jamming power, and hence, we obtain the closed-form solution of the optimal jamming power that maximizes the secrecy rate.
\end{abstract}


\section{Introduction}\label{sec:Intro}
\subsection{Background and Motivation}
\IEEEPARstart{D}{ue} to  the rapid adoption of wireless technologies in modern life, an unprecedented amount of private information is transmitted in wireless medium.  Consequently, communication security has become a critical issue due to the unalterable open nature of wireless channels.
As a complement to the traditional cryptographic technique, physical layer security has been extensively studied~\cite{Bloch_11,Zhou_13_Physical} to secure wireless communications by exploiting the characteristics of wireless channels.
Wyner introduced the wiretap-channel system as a framework for the physical layer security in his seminal work~\cite{wyner_75},  and defined the secrecy capacity as the maximum rate at which messages can be reliably sent to the intended receiver without being eavesdropped. This result was then generalized to the broadcast channel with confidential messages by Csisz\'{a}r and K\"{o}rner~\cite{csiszar_78} and the Gaussian wiretap channel by Leung-Yan-Cheong and Hellman~\cite{Cheong_78}.
In recent years, the fast development of multi-input multi-output (MIMO) techniques has triggered a considerable amount of attention on physical layer security in multi-antenna systems, where the transmitter, the receiver and/or the eavesdropper are equipped with multiple antennas. For example, the secrecy capacity of the multi-antenna system was analyzed in~\cite{Khisti_10,Khisti_10_II,Oggier_11_The} and signal processing techniques with multiple antennas for improving the secrecy performance were proposed in~\cite{Goel_08,Zhou_10,Zhou_11_secure,Mukherjee_11,Yang_13_Transmit}.

Despite a significant amount of work has been done on physical layer security, most research in this area is theoretically oriented due to the idealized and impractical assumptions.  For instance, many existing articles assumed that the transmitter has perfect channel state information (CSI) for the channels to the intended receiver and the eavesdropper.
In practice, an external eavesdropper naturally does not cooperate with the transmitter to send CSI feedback, and hence, it is very difficult for the transmitter to obtain the CSI of the eavesdropper. Although the intended receiver may cooperate to send CSI feedback, reliable uplink channels for the feedback cannot always be guaranteed.
This leads to an increasing amount of recent work focusing on the scenario where the transmitter does not have perfect CSI of the channel to the intended receiver and/or the eavesdropper, e.g., \cite{He_13_2,He_13_3,Cumanan_14_Secrecy,Chu_15_Secrecy,Chu_15_Robust} and references within.

Apart from the assumption of perfect CSI knowledge, another idealized assumption is often adopted in the existing literature on physical layer security in multi-antenna systems, i.e., the assumption of knowing the  number of eavesdropper antennas or setting an upper bound on the number of eavesdropper antennas.
If the number of eavesdropper antennas is unknown, we have to assume that the eavesdropper has an infinite number of antennas as a worst-case consideration, and then the secrecy rate would always go to zero intuitively.
To the best of the authors' knowledge, no existing literature has studied the scenario where the number of eavesdropper antennas is totally unknown.
In practice, an external eavesdropper naturally does not inform the legitimate side about the number of antennas to expose its ability.
As a weak justification, the upper bound on the number of eavesdropper antennas could be estimated from the eavesdropper's device size.
However, such a weak justification, probably valid in the past, can no longer hold with the current development of large-scale antenna array technologies which allow a fast growing number of antennas be placed within a limited space.
Thus, how to characterize the performance of
physical layer security without knowing the number of eavesdropper antennas is a challenging but important problem.

\subsection{Our Approach and Contributions}
In this work, we provide an innovative solution to the challenging problem by introducing the concept of spatial constraint\footnote{Here the spatial constraint
means the limited size of the spatial region for placing antennas at the communication node.} into physical layer security.
In practice, knowing the eavesdropper's spatial constraint for placing antennas is much easier than knowing the exact number of the eavesdropper antennas.
For example, we may know the size of the eavesdropper's device, but it is difficult to know how many antennas are installed on the device. We may know that the eavesdropper hides in a room, but it is difficult to known how many antennas are placed inside the room.
In addition, considering a spatial constraint instead of an exact location of the eavesdropper allows certain degrees of uncertainty in eavesdropper's location.

We focus on the effects of spatial constraints at the receiver side.
Specifically, we consider the scenario where
the transmitter has a large number of antennas without spatial constraint while
both the intended receiver and the eavesdropper have spatial constraints to place the receive antennas.
This is a valid assumption given less geometrical size restriction for the base station to place a large number of transmit antennas, while the size of receiving device in the downlink is often relatively small~\cite{Pollock_03_On}.
Importantly, the number of receive antennas at the eavesdropper may not be known. We consider a simple and practical CSI assumption that the
instantaneous CSI is known at the receiver end (the intended receiver and the eavesdropper) but not at
the transmitter.
Under these assumptions and considerations, we derive the secrecy capacity of the spatially-constrained multi-antenna system, and study the potential benefits brought by two widely-adopted friendly-jamming techniques.
The two friendly-jamming techniques studied are the basic jamming technique and the artificial noise (AN) jamming technique: the former degrades both the intended receiver and the eavesdropper's channels, while the latter degrades only the eavesdropper's channel but does not affect the intended receiver's channel.
We find that a non-zero secrecy capacity is achievable for the spatially-constrained system with the help of friendly-jamming signals, even if the number of eavesdropper antennas is unknown and considered to be infinity as a worst case.

The primary contributions of this paper are summarized as follows.
\begin{enumerate}
  \item We introduce spatial constraints into physical layer security. To this end, we propose a framework to study physical layer security in multi-antenna systems with spatial constraints at the receiver side (both the intended receiver and the eavesdropper). We derive the secrecy capacity, and analyze the impact of spatial constraints on the secrecy capacity.
  \item For the first time, our proposed framework allows one to analyze physical layer security without the knowledge of the number of eavesdropper antennas. It relaxes the requirement on the knowledge of eavesdropper from knowing the number of antennas to knowing the spatial constraint. We show that a non-zero secrecy capacity is achievable even if the eavesdropper is assumed to have an infinite number of antennas. This is easily achieved by applying the basic friendly-jamming technique where the jammer sends random noise signals.
  \item We further study the impact of jamming power on the secrecy capacity of the spatially-constrained jammer-assisted systems. For the basic jammer-assisted system, we find that the secrecy capacity does not monotonically increase with the jamming power, and we obtain the closed-form solution of the optimal jamming power that maximizes the secrecy capacity. The optimality of the obtained solution is confirmed by the numerical result.
\end{enumerate}

The remainder of this paper is organized as follows.
Section~\ref{sec:sysmod} describes system models for studying physical layer security with spatial constraints at the receiver side.
In Section~\ref{sec:SpaceCapacity}, we first give the secrecy capacity of the proposed systems with the knowledge of the number of eavesdropper antennas.
The important case of not knowing the number of eavesdropper antennas is studied in Section~\ref{sec:WorstCase}, where the eavesdropper's receiver is assumed to be noise free and allowed to have infinitely many antennas for the worst-case consideration.
Finally, Section~\ref{sec:Concl} concludes the paper and discusses possible future research directions.

Throughout the paper, we adopt the following notations:
Scalars, vectors and matrices are denoted by lowercase/uppercase letters, boldface lowercase letters and boldface uppercase letters, respectively,
the circularly symmetric complex Gaussian vector with mean $\boldsymbol{\mu}$ and covariance matrix $\mathbf{C}$  is denoted by $\mathcal{CN}\left(\boldsymbol{\mu},\mathbf{C}\right)$, $\mathbf{I}_N$ denotes the $N\times N$ identity matrix,
$()^H$ denotes the conjugate transpose of a vector or a matrix,
$\left|\mathbf{X}\right|$ denotes the determinant of matrix $\mathbf{X}$,
$\mathbb{E}\{\cdot\}$ denotes the expectation operator,
$\left\lceil\cdot\right\rceil$ and $\left\lfloor\cdot\right\rfloor$ denote the ceiling operator and the floor operator, respectively, $[x]^+=\max(x,0)$.

\section{System Model}\label{sec:sysmod}
In this paper, we study physical layer security in multi-antenna systems with spatial constraints at the receiver side.
We assume that all communication nodes are equipped with multiple antennas and there exist spatial constraints at both the intended receiver and the eavesdropper.
That is,  the intended receiver and the eavesdropper have limited sizes of spatial regions for placing the receive antennas.
To focus on the impact of spatial constraints at the receiver side, we adopt the following two assumptions as briefly mentioned in Section~\ref{sec:Intro}. Firstly, we assume that there is no spatial constraint at the transmitter side for placing transmit antennas. Secondly, we assume that the transmitter has a large number of transmit antennas, and hence the capacity of the channel from the transmitter to the receiver is mainly restricted by the receiver side.
Note that the number of antennas at the base station is often predicted to be in the hundreds for the next generation wireless systems~\cite{Larsson_14_Massive,Andrews_14_What}.
These two assumptions were often adopted in the literature investigating the impact of spatial constraints at the receiver side on multi-antenna systems without secrecy considerations, e.g.,~\cite{Pollock_03_J_Introducing,Pollock_03_On,Abhayapala_02_On,Pollock_03_bound} studying the channel capacity and~\cite{Kennedy_07_Instinsic,Bashar_12_Degrees,Bashar_14_Analysis,Bashar_14_Band} studying the spatial degrees of freedom.
We specifically investigate two different secure communication systems, which are the wiretap-channel system and the jammer-assisted system. For the jammer-assisted system, we further consider two different cases depending on the adopted jamming technique, namely basic jammer-assisted system and AN jammer-assisted system. The details of the system models are given in the following subsections.

\subsection{Wiretap-Channel System}\label{sec:Sys_M1}
\begin{figure}[!t]
\centering
\vspace{-0mm}
\includegraphics[width=0.7\columnwidth]{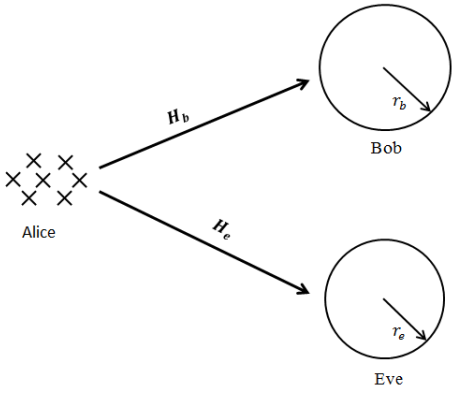}
\vspace{0mm}
\caption{2D model for the wiretap-channel system}
\vspace{-0mm}  \label{fig:M_TwoDOne}
\end{figure}

The wiretap-channel system consists of a transmitter, an intended receiver and an eavesdropper, with $N_t, N_b$ and $N_e$ antennas, respectively.
The transmitter, Alice, sends confidential messages to the intended receiver, Bob, in the presence of the eavesdropper, Eve.
The receive antennas at Bob and Eve are both spatially constrained.
Alice is assumed to be a base station with a large number of antennas $(N_t\rightarrow\infty)$ without a spatial constraint.
For the 2D analysis, Bob and Eve are assumed to be spatially constrained by circular apertures with radii $r_b$ and $r_e$, respectively.
For the 3D analysis, Bob and Eve are assumed to be spatially constrained by spherical apertures with radii $r_b$ and $r_e$, respectively.
The 2D model of the wiretap-channel system is depicted in Figure~\ref{fig:M_TwoDOne}. 

%


The received signal vector at Bob or Eve is given by
\begin{equation}\label{eq:yi_M1}
  \by_i=\sqrt{\alpha_i}\bH_i\bx+\bn_i  ~~i= b ~\text{or}~ e,
\end{equation}
where the subscripts $b$ and $e$ denote the parameters for Bob and
Eve, respectively, $\bx$ denotes the transmitted signal vector from Alice with an average power of $P_t$, i.e., $\mathbb{E}\left\{\bx^H\bx\right\}=P_t$.
In addition, $\bn_i\sim\mathcal{CN}\left(\mathbf{0},\sigma_i^2\bI\right)$ denotes the additive white Gaussian noise (AWGN) vector at Bob or Eve, $\bH_i=\left[\bh_{i1} \bh_{i2}\cdots\bh_{iN_t}\right]$ denotes the $N_i\times N_t$ normalized channel matrix from Alice to Bob or Eve with $\bh_{ik}$ $(k\in\{1, 2, \cdots, N_t\})$ representing the $N_i\times1$ complex zero-mean Gaussian vector of the channel gains corresponding to the $k$th transmit antenna at Alice. Moreover, $\alpha_i$ denotes the average channel gain from Alice to Bob or Eve, which is often determined by the distance between the transmitter and the receiver.
Besides, we assume that Bob and Eve perfectly know their CSI, while Alice does not know either Bob or Eve's instantaneous CSI.

The correlation matrix at the receiver is defined as
\begin{equation}\label{eq:RRh}
  \bR_i=\mathbb{E}\left\{\bh_{it}\bh_{it}^H\right\},  
\end{equation}
where the expectation is over all transmit antennas and channel realizations. We can also write
\begin{equation}\label{eq:RRrho}
  \bR_i=\left[
    \begin{array}{cccc}
      \rho_{i,11} &  \rho_{i,12} &  &  \rho_{i,1N_i} \\
       \rho_{i,21} &  \rho_{i,22} &  &  \\
      \vdots &  & \ddots & \vdots \\
       \rho_{i,N_i1} &  & \cdots &  \rho_{i,N_iN_i} \\
    \end{array}
  \right],
\end{equation}
with elements $\rho_{i,kk'}$ corresponding to the spatial correlation between two sensors $k$ and $k'$ at the receiver. The spatial correlation  between sensors is mainly determined by the distance between the sensors. The spatial correlation increases as the distance between sensors decreases.  Within a fixed space, the distance between the antennas decreases as the number of antennas increases, and hence, the spatial correlation increases as the number of antennas increases.

\subsection{Jammer-Assisted System}
\begin{figure}[!t]
\centering
\vspace{-0mm}
\includegraphics[width=0.7\columnwidth]{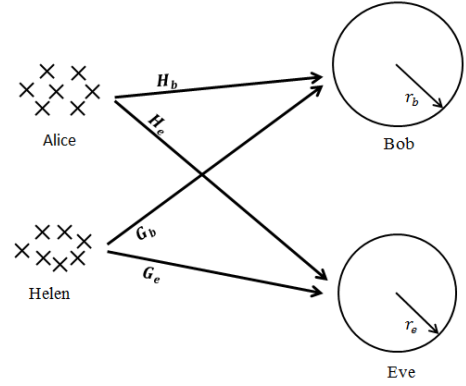}
\vspace{0mm}
\caption{2D model for the jammer-assisted system}
\vspace{-0mm}  \label{fig:M_TwoDTwo}
\end{figure}

The jammer-assisted system consists of a transmitter, a helper, an intended receiver and an eavesdropper, with $N_t, N_j, N_b$ and $N_e$ antennas, respectively.
With the aid of the helper, Helen, the transmitter, Alice, sends confidential messages to the intended receiver, Bob, in the presence of the eavesdropper, Eve.
Helen helps Alice by broadcasting friendly jamming signals.
The receive antennas at Bob and Eve are both spatially constrained.
Alice and Helen are assumed to be base stations with a large number of transmit antennas ($N_t, N_j\rightarrow\infty$) without the spatial constraint.
The detailed assumptions of the spatial constraints on Bob and Eve are the same as those given in Section~\ref{sec:Sys_M1}.
The 2D model of the jammer-assisted system is depicted in Figure~\ref{fig:M_TwoDTwo}. 

%

We assume that Bob and Eve perfectly know their CSI, and Alice does not know either Bob or Eve's instantaneous CSI.
We further assume that Helen does not know Eve's instantaneous CSI, since the passive eavesdropper does not feed back the CSI to the helper.
Moreover, for Helen's knowledge about Bob's channel, we consider two different cases in order to study two widely-adopted  friendly-jamming techniques, as will be detailed next.

\subsubsection{Case 1: Basic Jammer-Assisted System}\label{sec:Sys_M2C1}
In the first case, we assume that Helen does not know Bob's instantaneous CSI. This happens when there is no reliable uplink channel from Bob to Helen for CSI feedback. In this case, Helen broadcasts basic jamming signals that degrade both Bob and Eve's channels.

The received signal vector at Bob or Eve is given by
\begin{equation}\label{eq:yi_M2C1}
  \by_i=\sqrt{\alpha_i}\bH_i\bx+\sqrt{\beta_i}\bG_i\bw_1+\bn_i, ~~i= b ~\text{or}~ e,
\end{equation}
where $\bx, \alpha_i, \bH_i, \bn_i$ and the subscripts $b, e$ follow~\eqref{eq:yi_M1}. In addition, $\bw_1$ denotes the basic jamming signal vector transmitted from Helen with an average power of $P_j$, i.e., $\mathbb{E}\left\{\bw_1^H\bw_1\right\}=P_j$, and $\bG_i=\left[\bg_{i1} \bg_{i2}\cdots\bg_{iN_j}\right]$ denotes the normalized channel matrix from Helen to Bob or Eve  with $\bg_{ik}$ $(k\in\{1, 2, \cdots, N_j\})$ representing the $N_i\times1$ complex zero-mean Gaussian vector of the channel gains corresponding to the $k$th transmit antenna at Helen.
Moreover, $\beta_i$ denotes the average channel gain from Helen to Bob or Eve.


\subsubsection{Case 2: AN Jammer-Assisted System}\label{sec:Sys_M2C2}
In the second case, we assume that Helen perfectly knows the  instantaneous CSI from herself to Bob.
This happens when there exists a reliable uplink channel from Bob to Helen for CSI feedback. In such a case, Helen broadcasts AN jamming signals that degrade Eve's channel but do not affect Bob's channel.
The AN jamming technique was proposed in~\cite{Goel_08}, which is often applied in secure communication networks where the jammer has the CSI to the intended receiver.
Specifically, the AN jamming signal vector from Helen, denoted by $\bw_2$, is chosen to lie in the null space of the channel to the intended receiver, $\bG_b$. That is $\bG_b\bw_2=\mathbf{0}$.
In particular, $\bw_2$ can be constructed by
\begin{equation}\label{}
 \bw_2=\bZ\bv,
\end{equation}
where $\bv$ is an independent and identically distributed (i.i.d.) complex Gaussian random variable vector, the $N_j\times(N_j-N_b)$ matrix $\bZ$ denotes the orthonormal basis of the null space of $\bG_b$ with $\bZ^H\bZ=\bI$.

With the AN jamming signals, the received signal vectors at Bob and Eve are given by
\begin{equation}\label{eq:ybC3}
  \by_b=\sqrt{\alpha_b}\bH_b\bx+\sqrt{\beta_b}\bG_b\bw_2+\bn_b=\sqrt{\alpha_b}\bH_b\bx+\bn_b
\end{equation}
and
\begin{eqnarray}\label{eq:yeC3}
  \by_e  &=& \sqrt{\alpha_e}\bH_e\bx+\sqrt{\beta_e}\bG_e\bw_2+\bn_e \nonumber\\
   &=& \sqrt{\alpha_e}\bH_e\bx+\sqrt{\beta_e}\bG_e\bZ\bv+\bn_e,
\end{eqnarray}
respectively, where,  once again, $\bx, \alpha_b, \alpha_e, \bH_b, \bH_e, \bn_b,  \bn_e$  follow~\eqref{eq:yi_M1} and $\beta_b, \beta_e, \bG_b, \bG_e$ follow~\eqref{eq:yi_M2C1}. Besides, the average transmit power at Helen is still given by $P_j$, i.e., $\mathbb{E}\left\{\bw_2^H\bw_2\right\}=P_j$.

\begin{Remark}\label{Remark:Motivation_AN}
We highlight that the analysis for the AN jammer-assisted system is mainly motivated by its importance from the theoretical point of view. The basic jamming and the AN jamming are the two most widely-studied physical-layer techniques to improve the secrecy performance of multi-antenna systems.
In this work, we study the wireless physical layer security with spatial constraints at the receiver side. It is of significant importance to investigate the benefits brought by both of the jamming techniques in the spatially-constrained systems.
The AN jamming technique is often studied in the scenario where both Alice and
Helen have the legitimate CSI in the literature. The legitimate CSI available at Alice enables not only the injection of AN jamming signals but also the transmit beamforming, and the secrecy capacity will go to infinity under the assumption of infinitely large number of transmit antennas. This will be shown later in Section~\ref{sec:SpaceCapacity}.
In order to investigate the capacity improvement solely brought by AN jamming, we assume that Alice does not know the instantaneous CSI to Bob, but Helen knows the instantaneous CSI to Bob. Besides, the practical value of the AN jammer-assisted system studied in this paper can be seen from the following scenario as an example:
We can consider that Alice is a base station owned by company A to serve a mobile user, Bob. Helen is another base station owned by company B.
Due to particular reasons, e.g., location or surrounding environment, the CSI feedback link from Bob to Alice is bad, while the CSI feedback link from Bob to Helen is good.
Then, Alice asks Helen to help the secrecy transmission by broadcasting AN jamming signals. For the secrecy concern, company A does not intend to share the confidential information with company B, and hence
Alice does not share the messages to transmit with Helen.
\end{Remark}

%

\section{Introducing Spatial Constraints into Secrecy Capacity Calculation}\label{sec:SpaceCapacity}
In this section, we derive the secrecy capacity of the systems with spatial constraints at the receiver side as described in Section~\ref{sec:sysmod}. The secrecy capacity characterizes the maximum rate at which messages can be reliably transmitted to Bob while Eve obtains zero information. It is mathematically defined by~\cite{Cheong_78}
\begin{equation}\label{eq:CsB}
  C_s=[C_b-C_e]^+,
\end{equation}
where $C_b$ and $C_e$ denote Bob and Eve's channel capacities, respectively.

For the multi-antenna systems with spatial constraint at the receiver, the channel capacity is limited by the rank and the eigenvalues of the spatial correlation matrix at the receiver. As the number of antennas increases in a fixed space, the correlation between antennas increases. The increase in spatial correlation will limit the number of significant eigenvalues of the spatial correlation matrix. As more antennas are placed in the fixed space, they will be highly correlated with other antennas. As a result, the growth of channel capacity with respect to the number of receive antennas reduces from linear to logarithmic.
The number of receive antennas at which the capacity scaling is reduced to logarithmic is approximated by the saturation number of receive antennas. The saturation number of receive antennas is given by~\cite[Chapters~3.3]{Pollock_03_On}
\begin{equation}\label{eq:Nsati}
  \Nsati=\left\{ \begin{array}{ll}
  2\left\lceil\pi e r_i/\lam \right\rceil+1\;, &\mbox{for 2D analysis}     \\
  (\left\lceil\pi e r_i/\lam \right\rceil+1)^2\;, &\mbox{for 3D analysis},
  \end{array}
  \right.
\end{equation}
where $\lam$ denotes the wavelength, $e$ denotes Euler's number, and subscript $i$ denotes the parameters for Bob or Eve.\footnote{The detailed derivation of the saturation number of antennas closely follows~\cite[Chapters~2.1 and 3]{Pollock_03_On}.}
As pointed out in~\cite{Pollock_03_On}, the growth of channel capacity ($C_b$ or $C_e$) with respect to the number of \emph{optimally-placed} receive antennas ($N_b$ or $N_e$) reduces from linear to logarithmic when the number of receive antennas increases beyond the saturation number ($\Nsatb$ or $\Nsate$).
Note that similar ``saturation" effects on the growth of channel capacity with respect to the number of antennas at the spatially-constrained receiver have also been pointed out in, e.g.,~\cite{Gesbert_02_Capacity,Hanlen_02_Capacity,Pollock_03_Antenna,Wu_07_On}.

It is worth mentioning that the capacity results in this paper are approximations based on \eqref{eq:Nsati} and the assumption of infinitely large number of transmit antennas. The accuracy of the approximations are verified in Appendices. In the rest of the paper, we simply refer to the approximated capacity result as the capacity.
\subsection{Secrecy Capacity of Wiretap-Channel System}
\begin{Proposition}\label{Pro:CsM1}
The secrecy capacity of the wiretap-channel system with spatial constraints at the receiver side is given by $C_s=[C_b-C_e]^+$ where\footnote{Throughout the paper logarithms are to base two, and the capacity is therefore in bits/s/Hz.}
\begin{equation}\label{}
  C_b = \left\{ \begin{array}{ll} N_b\log(1+\frac{\alpha_bP_t}{\sgs_b})\;, &\text{if}~N_b\le\Nsatb\\
  \Nsatb\log(1+\frac{N_b}{\Nsatb}\frac{\alpha_bP_t}{\sgs_b})\;, & \text{otherwise},
  \end{array}
  \right.
\end{equation}
\begin{equation}\label{}
  C_e = \left\{ \begin{array}{ll} N_e\log(1+\frac{\alpha_eP_t}{\sgs_e})\;, &\text{if}~N_e\le\Nsate\\
  \Nsate\log(1+\frac{N_e}{\Nsate}\frac{\alpha_eP_t}{\sgs_e})\;, & \text{otherwise.}
  \end{array}
  \right.
\end{equation}
\begin{IEEEproof}
The capacities of the channels to the spatially-constrained Bob and Eve follow easily from~~\cite[Chapters~2 and 3]{Pollock_03_On}. The details are given in Appendix~\ref{sec:ProofPreli}.
\end{IEEEproof}
\end{Proposition}

\textit{Proposition~\ref{Pro:CsM1}} gives the secrecy capacity of the wiretap-channel system taking spatial constraints at the receiver side into account.
From Proposition~\ref{Pro:CsM1}, we note that the growth of secrecy capacity with $N_b$ reduces from linear to logarithmic once $N_b$ reaches $\Nsatb$.
Also, the decrease of secrecy capacity with $N_e$ reduces from linear to logarithmic once  $N_e$ reaches $\Nsate$.
Differently, the secrecy capacity without spatial constraint always increases linearly with $N_b$ and decreases linearly with $N_e$.
This verifies that the secrecy performances of the networks with and without spatial considerations are different.

\subsection{Secrecy Capacity of Basic Jammer-Assisted System}\label{sec:Cs_M2C1}
\begin{Theorem}\label{Theo:CsM2C1}
The secrecy capacity of the basic jammer-assisted system with spatial constraints at the receiver side is given by $C_s=[C_b-C_e]^+$ where
\begin{equation}\label{}
C_b = \left\{ \begin{array}{ll} N_b\log\left(1+\frac{\alpha_bP_t}{\beta_bP_j+\sgs_b}\right)\;, &\text{if}~N_b\le\Nsatb\\
  \Nsatb\log\left(1+\frac{\frac{N_b}{\Nsatb}\alpha_bP_t}{\frac{N_b}{\Nsatb}\beta_bP_j+\sgs_b}\right)\;, & \text{otherwise},
  \end{array}
  \right.\!\!
\end{equation}
\begin{equation}\label{}
C_e = \left\{ \begin{array}{ll} N_e\log\left(1+\frac{\alpha_eP_t}{\beta_eP_j+\sgs_e}\right)\;, &\text{if}~N_e\le\Nsate\\
  \Nsate\log\left(1+\frac{\frac{N_e}{\Nsate}\alpha_eP_t}{\frac{N_e}{\Nsate}\beta_eP_j+\sgs_e}\right)  \;, & \text{otherwise.}
  \end{array}
  \right.\!\!
\end{equation}
%
%
%
\begin{IEEEproof}
See Appendix~\ref{sec:ProofCsM2C1}.
\end{IEEEproof}
\end{Theorem}

\textit{Theorem~\ref{Theo:CsM2C1}} gives the secrecy capacity of the basic jammer-assisted system taking spatial constraints at the receiver side into account. Similar to the result for the wiretap channel, we note that the secrecy capacity grows in linear with $N_b$ when $N_b\le\Nsatb$. Also, the secrecy capacity decreases in linear with $N_e$ when $N_e\le\Nsate$.
However, as $N_i$ increases beyond $\Nsati$, the change of secrecy capacity with respect to $N_i$ becomes slower and slower. The secrecy capacity approaches an upper bound
as $N_b\rightarrow\infty$, and a (possible) non-zero lower bound as  $N_e\rightarrow\infty$, since
\begin{equation}\label{}
  \lim_{N_b\rightarrow\infty}C_b=\Nsatb\log\left(1+\frac{\alpha_bP_t}{\beta_bP_j}\right)
\end{equation}
and
\begin{equation}\label{}
   \lim_{N_e\rightarrow\infty}C_e=\Nsate\log\left(1+\frac{\alpha_eP_t}{\beta_eP_j}\right).
\end{equation}

\subsection{Secrecy Capacity of AN Jammer-Assisted System}
\begin{Theorem}\label{Theo:CsM2C2}
The secrecy capacity of the AN jammer-assisted system with spatial constraints at the receiver side is given by $C_s=[C_b-C_e]^+$ where
\begin{equation}\label{}
  C_b = \left\{ \begin{array}{ll} N_b\log\left(1+\frac{\alpha_bP_t}{\sgs_b}\right)\;, &\text{if}~N_b\le\Nsatb\\
 \Nsatb\log(1+\frac{N_b}{\Nsatb}\frac{\alpha_bP_t}{\sgs_b})\;, & \text{otherwise},
  \end{array}
  \right.
\end{equation}
\begin{equation}\label{eq:CeJa}
  C_e = \left\{ \begin{array}{ll} N_e\log\left(1+\frac{\alpha_eP_t}{\beta_eP_j+\sgs_e}\right)\;, &\text{if}~N_e\le\Nsate\\
  \Nsate\log\left(1+\frac{\frac{N_e}{\Nsate}\alpha_eP_t}{\frac{N_e}{\Nsate}\beta_eP_j+\sgs_e}\right)\;, & \text{otherwise.}
  \end{array}
  \right.\!\!
\end{equation}
\begin{IEEEproof}
The capacity of Bob's channel is the same as that for the wiretap-channel system, since the AN jamming signals do not affect Bob's channel. We then derive the capacity of Eve's channel subject to the AN jamming signals.
The details are given in Appendix~\ref{sec:ProofCsM2C2}.
\end{IEEEproof}
\end{Theorem}

\textit{Theorem~\ref{Theo:CsM2C2}} gives the secrecy capacity of the AN jammer-assisted system taking spatial constraints at the receiver side into account.
We note that the growth of secrecy capacity with $N_b$ reduces from linear to logarithmic once $N_b$ reaches $\Nsatb$. The decrease of secrecy capacity with $N_e$ is in linear when $N_e\le\Nsate$, and becomes slower and slower when $N_e>\Nsate$. The secrecy capacity approaches a (possible) non-zero lower bound as $N_e\rightarrow\infty$.

\subsection{Secrecy Capacity with Legitimate CSI Available at Alice}\label{sec:CSIavailableatAlice}
In this paper, we consider a simple and practical CSI assumption that the
instantaneous CSI of Bob is not available at Alice. In fact, it is also possible in practice that Bob's CSI is available at Alice. In this subsection, we provide the analysis on the secrecy capacity of the scenario where both Alice and Helen have Bob's CSI.
Note that for the scenario without the friendly jammer, Alice can use a portion of the transmit antennas for sending information signals and the rest for broadcasting AN jamming signals.
Under the assumption of $N_t\rightarrow\infty$, the scenario without the jammer Helen can be regarded as the scenario having both Helen and Alice at the same location.

When Bob's CSI is available at Alice, Alice can design the transmit signals accordingly to enhance Bob's channel capacity. At the same time, Helen can still transmit the AN jamming signals that degrade Eve's channel but do not affect Bob's channel.
An infinitely large rate at Bob can be achieved by adopting a simple single-stream beamforming at Alice, under the assumption that the transmitter has an infinitely large number of antennas without the spatial constraint, while Eve does not benefit from the transmit beamforming. Hence, the secrecy capacity is equal to infinity in such a scenario with Bob's CSI available at Alice.
It is worth mentioning that the secrecy capacity would be finite in a practical system with spatial constraints at both the transmitter side and the receiver side, due to the finite degrees of freedom in the spatially-constraint channel.
The derivation of secrecy capacity in systems with spatial constraints at both the transmitter side and the receiver side is non-trivial and beyond the scope of this paper.

\subsection{Numerical Results}
In this subsection, we demonstrate the secrecy capacity versus the number of Bob's antennas and the number of Eve's antennas for different systems.
Specifically, the network parameters are
$P_t=20~\mbox{dB}, P_j=0~\mbox{dB}, \alpha_b=1, \alpha_e=1, \beta_b=1, \beta_e=1, \sigma_b^2=1, \sigma_e^2=1,  r_b=1.5\lambda~\mbox{and}~r_e=1\lambda$.
We adopt the 2D analysis to characterize the spatial constraints at the receiver side. That is, Bob and Eve are assumed to be spatially constrained by circular apertures. According to \eqref{eq:Nsati}, the saturation numbers of receive antennas  for Bob and Eve are $\Nsatb=2\left\lceil\pi e r_b/\lam \right\rceil+1=27$ and $\Nsate=2\left\lceil\pi e r_e/\lam \right\rceil+1=19$, respectively.

%

\begin{figure}[!t]
\centering
\vspace{-0mm}
\includegraphics[height=2.8in,width=3.3in]{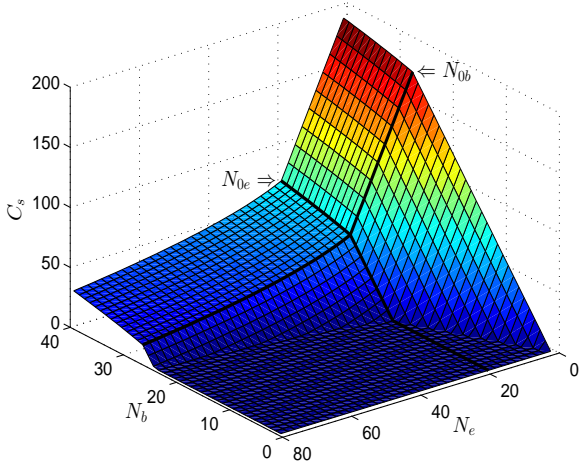}
\vspace{-0mm}
\caption{Wiretap-channel system: Secrecy capacity versus the number of Bob's antennas and the number of Eve's antennas. Bob and Eve are spatially constrained by circular apertures with  radii $r_b=1.5\lambda~\mbox{and}~r_e=1\lambda$, respectively.}
\vspace{-0mm}  \label{fig:Cs3D1}
\end{figure}

\begin{figure}[!t]
\centering
\vspace{-0mm}
\includegraphics[height=2.8in,width=3.3in]{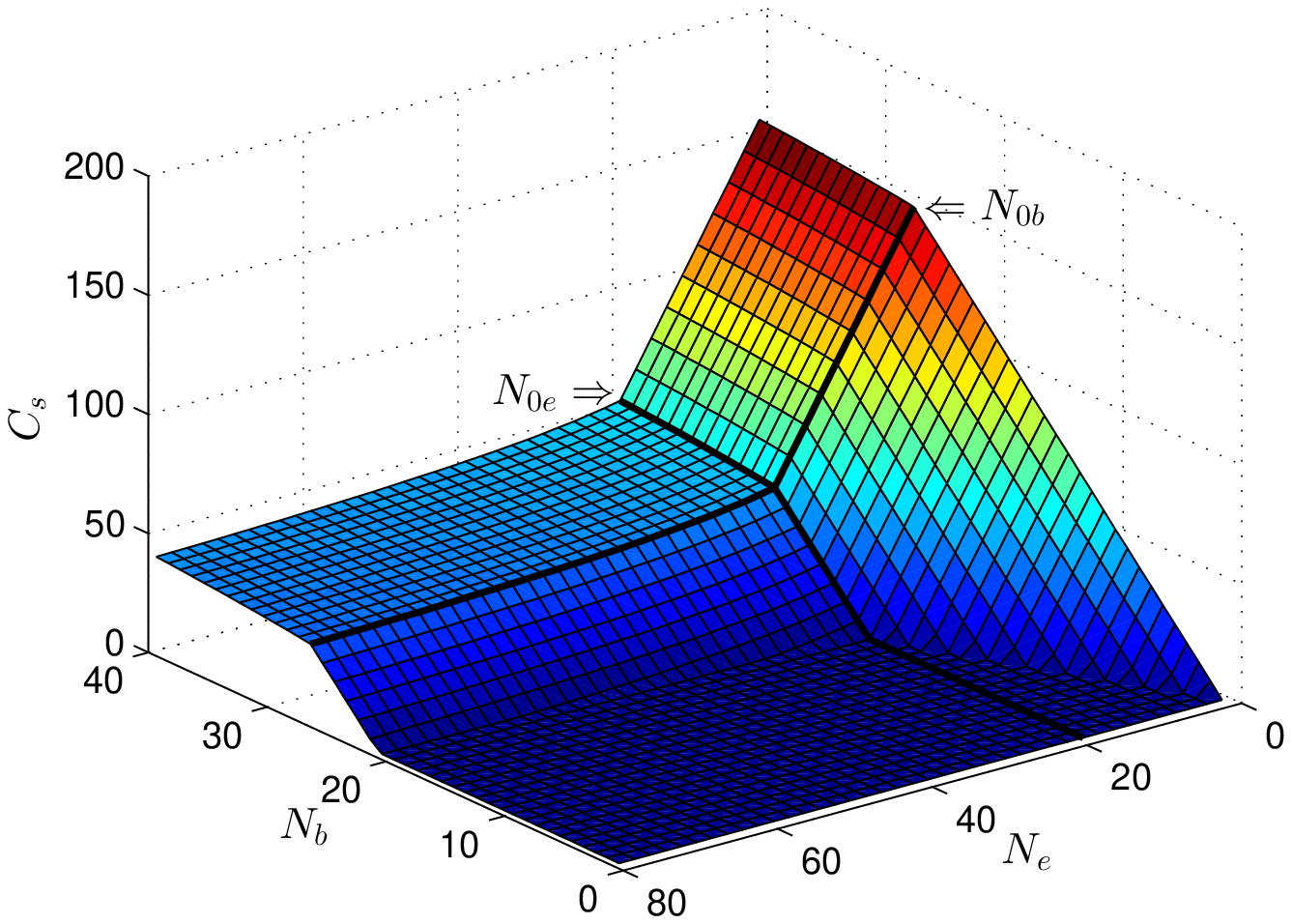}
\vspace{-0mm}
\caption{Basic jammer-assisted system: Secrecy capacity versus the number of Bob's antennas and the number of Eve's antennas.  Bob and Eve are spatially constrained by circular apertures with  radii $r_b=1.5\lambda~\mbox{and}~r_e=1\lambda$, respectively.}
\vspace{-0mm}  \label{fig:Cs3D2}
\end{figure}

\begin{figure}[!t]
\centering
\vspace{-0mm}
\includegraphics[height=2.8in,width=3.3in]{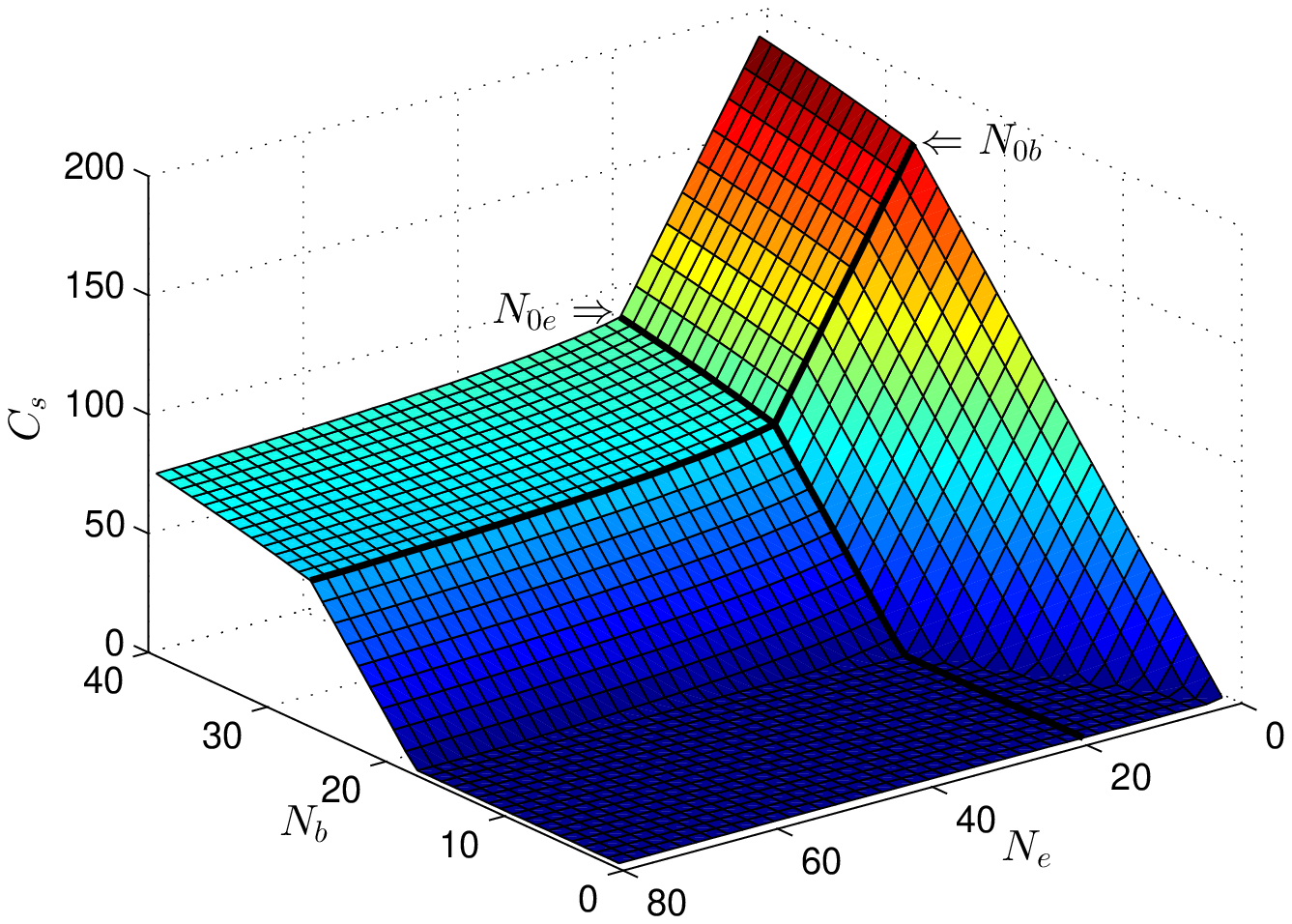}
\vspace{-0mm}
\caption{AN jammer-assisted system: Secrecy capacity versus the  number of Bob's antennas and the number of Eve's antennas.  Bob and Eve are spatially constrained by circular apertures with  radii $r_b=1.5\lambda~\mbox{and}~r_e=1\lambda$, respectively.}
\vspace{-0mm}  \label{fig:Cs3D3}
\end{figure}

\begin{figure}[!t]
\centering
\vspace{0mm}
\includegraphics[height=2.8in,width=3.3in]{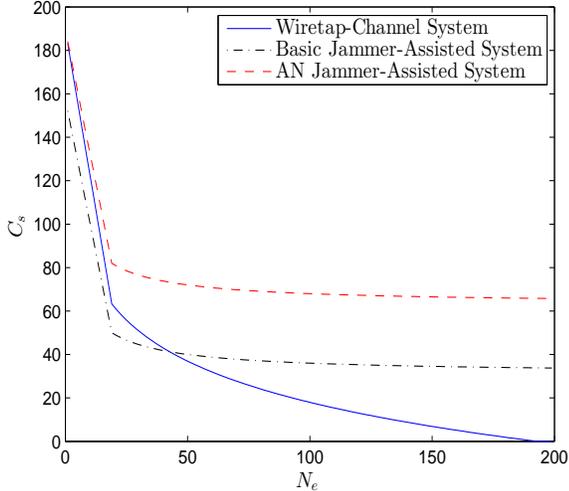}
\vspace{-0mm}
\caption{Secrecy capacity versus the  number of eavesdropper antennas with $N_b=35$. Bob and Eve are spatially constrained by circular apertures with radii $r_b=1.5\lambda~\mbox{and}~r_e=1\lambda$, respectively.}
\vspace{-0mm}  \label{fig:Cs2D123}
\end{figure}

Figures \ref{fig:Cs3D1}, \ref{fig:Cs3D2} and \ref{fig:Cs3D3} plot $C_s$ versus $N_b$ and $N_e$ for the wiretap-channel system, the basic jammer-assisted system and the AN jammer-assisted system, respectively.
As shown in the figures, $C_s$ increases with $N_b$ and decreases with $N_e$. The increase of $C_s$ with $N_b$ slows down once $N_b>\Nsatb$ due to the effect of spatial constraint at Bob. Similarly, the decrease of $C_s$ with $N_e$ slows down once $N_e>\Nsate$ due to the effect of spatial constraint at Eve. Besides, we note that the achieved secrecy capacities for different systems are different.

To make a clear comparison between the achieved secrecy capacities for different systems, we present Figure~\ref{fig:Cs2D123} plotting $C_s$ versus $N_e$ with a given value of $N_b=35$.
As shown in the figure, the secrecy capacity of the wiretap-channel system decreases fast as the number of Eve's antennas increases. 
We find that the secrecy capacity of the wiretap-channel system goes to zero as the number of Eve's antennas continues to increase.
Comparing the wiretap-channel system and the basic jammer-assisted system, we note that introducing the basic jamming signals effectively slows down the decrease of $C_s$ when $N_e>\Nsate$.
Thus, the basic jammer-assisted system achieves a higher secrecy capacity compared with the wiretap-channel system when the number of Eve's antennas is large.
In addition, as analyzed in Section~\ref{sec:Cs_M2C1}, the secrecy capacity of the basic jammer-assisted system can approach a non-zero lower bound as $N_e\rightarrow\infty$.
Besides, we observe from the figure that the secrecy capacity achieved by the basic jammer-assisted system is less than that achieved by the wiretap-channel system when $N_e$ is small. This is actually because the jamming power has not been optimally designed for the results in the figure.
Comparing the wiretap-channel system and the AN jammer-assisted system, we find that the AN jammer-assisted system always obtains a higher secrecy capacity than that of the wiretap-channel system. This is because the AN jamming signals degrade Eve's channel only, but do not affect Bob's channel. However, we should note that broadcasting the AN jamming signals requires the helper to know the instantaneous CSI of the intended receiver, which is not always possible in practice.

\section{Worst-Case Analysis for Jammer-Assisted Systems}\label{sec:WorstCase}
The previous section provides the basic analysis on the secure communication systems with spatial constraints at the receiver side. However, to evaluate the system performance by the capacity results given in \textit{Proposition~\ref{Pro:CsM1}}, \textit{Theorems~\ref{Theo:CsM2C1}}  and \textit{\ref{Theo:CsM2C2}}, we require very good knowledge on Eve, including $N_e$ and $\sgs_e$.
In practice, it is desirable to be able to investigate the secrecy performance of a system without the knowledge of $N_e$ and $\sgs_e$. To this end, we consider a ``worst-case eavesdropper" (from the legitimate users' perspective) as in this section.

For such a worst-case eavesdropper, we assume that the number of receive antennas at the eavesdropper approaches infinity and the noise variance at the eavesdropper approaches zero, i.e, $N_e\rightarrow\infty~\text{and}~\sgs_e\rightarrow0.$
Then, the secrecy capacity with the worst-case consideration is given by
\begin{equation}\label{eq:Cs_Worst}
  C_s^\wor=\lim_{N_e\rightarrow\infty,\sgs_e\rightarrow0}C_s,
\end{equation}
where $C_s$ is the secrecy capacity of systems with perfect knowledge of $N_e$ and $\sgs_e$, i.e, the secrecy capacity derived in the previous section.
In addition, we refer to $C_s^\wor$ as the worst-case secrecy capacity.

The worst-case scenario is motivated by the fact that the eavesdropper's ability is difficult to be known or controlled by the legitimate side. As such, in the design of secure communications, we assume the worst-case scenario where the eavesdropper can deploy infinite number of antennas with arbitrarily small noise variance. If we assume that the eavesdropper has a given number of antennas, the designed secure communications would be vulnerable to eavesdropping caused by a larger number of antennas at the eavesdropper in practice. Therefore, the weaker assumption of knowing a finite number of antennas at the eavesdropper cannot lead to the true guarantee of security, and thus it is of critical significance to take into consideration the worst-case scenario with infinite number of eavesdropper antennas.



\subsection{Wiretap-Channel System}
Based on \textit{Proposition~\ref{Pro:CsM1}} and~\eqref{eq:Cs_Worst}, the worst-case secrecy capacity of the wiretap-channel system is given by
\begin{equation}\label{eq:Cs_WorstM1}
  C^\wor_{s}
  =0.
\end{equation}
We note that a non-zero worst-case secrecy capacity is not achievable under any condition for the wiretap-channel system, because the capacity of Eve's channel always goes to infinity with $N_e\rightarrow\infty~\text{or}~\sgs_e\rightarrow0$.

\subsection{Basic Jammer-Assisted System}
\subsubsection{Worst-Case Secrecy Capacity}
Based on \textit{Theorem~\ref{Theo:CsM2C1}} and~\eqref{eq:Cs_Worst}, the worst-case secrecy capacity of the basic jammer-assisted system is given by
\begin{equation}\label{eq:Cs_WorstM2C1}
  C^\wor_{s}
  = \left\{ \begin{array}{ll}
  \left[N_b\log\left(1+\frac{\alpha_bP_t}{\beta_bP_j+\sgs_b}\right)
 \!-\!\Nsate\log\left(1+\frac{\alpha_eP_t}{\beta_eP_j}\right)\right]^+\!\!\!,\;
  \\\text{if}~N_b\le\Nsatb \\
  \left[\Nsatb\log\left(1+\frac{\frac{N_b}{\Nsatb}\alpha_bP_t}{\frac{N_b}{\Nsatb}\beta_bP_j+\sgs_b}\right)
 \!-\!\Nsate\log\left(1+\frac{\alpha_eP_t}{\beta_eP_j}\right)\right]^+\!\!\!,\;
  \\\text{otherwise.}
  \end{array}
  \right.
\end{equation}
From~\eqref{eq:Cs_WorstM2C1}, we note that a non-zero worst-case secrecy capacity sometimes is achievable for the basic jammer-assisted system depending on the system parameters, such as transmit power, average channel gains, the spatial constraint at Bob and the number of antennas at Bob. 
This result shows for the first time that a non-zero secrecy rate can be achieved even if the eavesdropper's receiver itself is noise free and allowed to have infinitely many antennas. Moreover, this is achieved by simply asking a friendly-jamming node to send random jamming signals.

To further study the condition for having a non-zero worst-case secrecy capacity, we consider the scenario where the number of antennas at Bob, $N_b$, is controllable and the other system parameters\footnote{Here the other system parameters depend on the spatial constraint, the location of communication node and the transmit power.}, i.e., $\Nsatb, \Nsate, \alpha_b, \beta_b, \alpha_e, \beta_e, P_t$ and $P_j$, are fixed.
From~\eqref{eq:Cs_WorstM2C1}, we find that a non-zero worst-case secrecy capacity is always achievable by having ``enough" receive antennas at Bob when $\Nsatb\log\left(1+\frac{\alpha_bP_t}{\beta_bP_j}\right)>\Nsate\log\left(1+\frac{\alpha_eP_t}{\beta_eP_j}\right)$. However, the secrecy capacity is always equal to zero when
\begin{equation}\label{eq:NbinfinityCs0}
  \Nsatb\log\left(1+\frac{\alpha_bP_t}{\beta_bP_j}\right)\le\Nsate\log\left(1+\frac{\alpha_eP_t}{\beta_eP_j}\right),
\end{equation}
because $C_b<\Nsatb\log\left(1+\frac{\alpha_bP_t}{\beta_bP_j}\right)$
always holds for any finite value of $N_b$.
In addition, when $\Nsatb\log\left(1+\frac{\alpha_bP_t}{\beta_bP_j}\right)>\Nsate\log\left(1+\frac{\alpha_eP_t}{\beta_eP_j}\right)$,
we can further derive the minimum $N_b$ to ensure a non-zero worst-case secrecy capacity as
\begin{equation}\label{eq:Nbm_basic}
  \Nbm=\left\{ \begin{array}{ll}
  \left\lfloor\frac{\Nsate\log\left(1+\frac{\alpha_eP_t}{\beta_eP_j}\right)}{\log\left(1+\frac{\alpha_bP_t}{\beta_bP_j+\sgs_b}\right)}\right\rfloor+1\;, \\\text{if}~\Nsatb\log\left(1+\frac{\alpha_bP_t}{\beta_bP_j+\sgs_b}\right)
 \ge\Nsate\log\left(1+\frac{\alpha_eP_t}{\beta_eP_j}\right)  \vspace{1mm}\\
  \left\lfloor\frac{\mathrm{\Nsatb}\, {\mathrm{\sigma}_{b}}^2\, \left({\left(1+\frac{\mathrm{\alpha}_{e}P_{t}}{\mathrm{\beta}_{e}P_{j}} \right)}^{\frac{\Nsate}{\mathrm{\Nsatb}}} - 1\right)}
{\mathrm{\alpha}_{b}P_{t} + \mathrm{\beta}_{b}P_{j} - \mathrm{\beta}_{b}P_{j}
{\left(1+\frac{\mathrm{\alpha}_{e}P_{t}}{\mathrm{\beta}_{e}P_{j} }\right)}^{\frac{\Nsate}{\mathrm{\Nsatb}}}}\right\rfloor+1\;,
   \\\text{otherwise.}
  \end{array}
  \right.
\end{equation}

\subsubsection{Optimal Jamming Power}
From~\eqref{eq:Cs_WorstM2C1}, we note that the worst-case secrecy capacity is not a monotonically increasing function of the jamming power. This is because the increase of $P_j$ degrades not only Eve's channel but also Bob's channel, and there arises a tradeoff between maintaining the capacity of Bob's channel and decreasing the capacity of Eve's channel. In the following, we determine the optimal jamming power that maximizes the worst-case secrecy capacity, i.e.,  $P_j^\opt=\arg\max_{P_j} C^\wor_{s}$.

\begin{Proposition}\label{Pro:Opt_Pj}
The optimal jamming power that maximizes the worst-case secrecy capacity of the basic jammer-assisted system is given by
\begin{equation}\label{}
 P_j^\opt\!=\! \left\{ \begin{array}{llll}
  \!\!\!x_1,\; &\!\!\!\!\!\!\!\text{if}~N_b\le\Nsatb,\!~f_1(x_1)>0,\!~x_1~\text{is real and positive}\\
\!\!\!x_2,\; &\!\!\!\!\!\!\!\text{if}~N_b\le\Nsatb,\!~f_2(x_2)>0,\!~x_2~\text{is real and positive}\\
\!\!\!x_3,\; &\!\!\!\!\!\!\!\text{if}~N_b\le\Nsatb,\!~f_3(x_3)>0,\!~x_3~\text{is real and positive}\\
\!\!\!x_4,\; &\!\!\!\!\!\!\!\text{if}~N_b\le\Nsatb,\!~f_4(x_4)>0,\!~x_4~\text{is real and positive}\\
  \text{n/a}\;,  &\text{otherwise,} 
  \end{array}
  \right.
\end{equation}
where
\begin{equation*}
  f_1(x)= N_b\log\left(1+\frac{\alpha_bP_t}{\beta_bx+\sgs_b}\right)
 -\Nsate\log\left(1+\frac{\alpha_eP_t}{\beta_ex}\right),
\end{equation*}
\begin{equation*}
  f_2(x)=\Nsatb\log\left(1+\frac{\frac{N_b}{\Nsatb}\alpha_bP_t}{\frac{N_b}{\Nsatb}\beta_bx+\sgs_b}\right)
 - \Nsate\log\left(1+\frac{\alpha_eP_t}{\beta_ex}\right),
\end{equation*}

\begin{eqnarray*}\label{}
x_1&=&
\frac{  2\Nsate\alpha_{e}\sigma^2_{b}
  -P_{t}\alpha_{b}\alpha_{e}\left(N_b-\Nsate\right)
}
{
2\left(
  N_{b}\alpha_{b}\beta_{e}
  -  \Nsate\alpha_{e}\beta_{b}
  \right)
}\\
&&+
\frac{
  \sqrt{
  \alpha_{b}^2\alpha_{e}^2\beta_{b}^2\, P_{t}^2
  \left(N_{b}-\Nsate
  \right)^2
  +\phi_1}
}
{
2\beta_{b}
  \left(
  N_{b}\alpha_{b}\beta_{e}
  -  \Nsate\alpha_{e}\beta_{b}
  \right)
},
\end{eqnarray*}

\begin{eqnarray*}\label{}
x_2&=&
\frac{2\Nsate\alpha_{e}\sigma^2_{b}
  -P_{t}\alpha_{b}\alpha_{e}\left(N_b-\Nsate\right)
}
{
2\left(
  N_{b}\alpha_{b}\beta_{e}
  -  \Nsate\alpha_{e}\beta_{b}
  \right)
}\\
&&-
\frac{
  \sqrt{
  \alpha_{b}^2\alpha_{e}^2\beta_{b}^2\, P_{t}^2
  \left(N_{b}-\Nsate
  \right)^2
  +\phi_1
  }
}
{
2\beta_{b}
  \left(
  N_{b}\alpha_{b}\beta_{e}
  -  \Nsate\alpha_{e}\beta_{b}
  \right)
},
 \end{eqnarray*}

\begin{eqnarray*}\label{}
x_3&=&
\frac{
2\Nsate\Nsatb\alpha_{e}\sigma^2_{b}
-N_{b}P_{t}\alpha_{b}\alpha_{e}\left(\Nsatb-\Nsate\right)
}
{
2N_{b}
\left(
\Nsatb\alpha_{b}\beta_{e}
-\Nsate\alpha_{e}\beta_{b}
\right)
}\\
&&+
\frac{
\sqrt{\alpha_b^2\alpha_e^2\beta_b^2P_t^2N_b^2\left(\Nsatb-\Nsate\right)^2
+\phi_2
}
}
{2N_{b}\beta_{b}
\left(
\Nsatb\alpha_{b}\beta_{e}
-\Nsate\alpha_{e}\beta_{b}
\right)},
\end{eqnarray*}

\begin{eqnarray*}\label{}
x_4&=&
\frac{
2\Nsate\Nsatb\alpha_{e}\sigma^2_{b}
-N_{b}P_{t}\alpha_{b}\alpha_{e}\left(\Nsatb-\Nsate\right)
}
{
2N_{b}
\left(
\Nsatb\alpha_{b}\beta_{e}
-\Nsate\alpha_{e}\beta_{b}
\right)
}\\
&&-
\frac{
\sqrt{\alpha_b^2\alpha_e^2\beta_b^2P_t^2N_b^2\left(\Nsatb-\Nsate\right)^2
+\phi_2
}
}
{2N_{b}\beta_{b}
\left(
\Nsatb\alpha_{b}\beta_{e}
-\Nsate\alpha_{e}\beta_{b}
\right)},
\end{eqnarray*}
with
\begin{equation*}\label{}
  \phi_1=4N_{b}\Nsate\alpha_{b}\alpha_{e}\beta_{b}\sigma^2_{b}
  \left(
  P_{t}\alpha_{b}\beta_{e}
   - P_{t}\alpha_{e}\beta_{b}
  + \beta_{e}\sigma^2_{b}
  \right),
\end{equation*}
\begin{equation*}\label{} \phi_2\!=\!4\Nsatb^2\Nsate\alpha_b\alpha_e\beta_b\sigma_b^2\left(N_bP_t\alpha_b\beta_e\!-\!N_bP_t\alpha_e\beta_b\!+\!\Nsatb\beta_e\sigma_b^2\right)\!.
\end{equation*}

\end{Proposition}

\begin{IEEEproof}
See Appendix~\ref{sec:ProofOptPj}.
\end{IEEEproof}

\begin{Remark}
\textit{Proposition~\ref{Pro:Opt_Pj}} provides the optimal jamming power that maximizes the worst-case secrecy capacity of the basic jammer-assisted system.
If there is no power constraint at the jammer, we can simply set the jamming power as $ P_j^\opt$ to achieve the best secrecy performance. If there exists a power constraint at the jammer, say $P_j\le P_{j,\max}$, we should first check the feasibility of achieving the non-zero worst-case secrecy capacity, and then set the jamming power as $\min(P_j^\opt,P_{j,\max})$ if the non-zero worst-case secrecy capacity is achievable.
\end{Remark}

\subsection{AN Jammer-Assisted System}
Based on \textit{Theorem~\ref{Theo:CsM2C2}} and~\eqref{eq:Cs_Worst}, the worst-case secrecy capacity of the AN jammer-assisted system  is given by
\begin{equation}\label{eq:Cs_WorstM2C2}
  C^\wor_{s}
  = \left\{ \begin{array}{ll}
  \left[N_b\log\left(1+\frac{\alpha_bP_t}{\sgs_b}\right)
 -\Nsate\log\left(1+\frac{\alpha_eP_t}{\beta_eP_j}\right)\right]^+\;,
  \\ \text{if}~N_b\le\Nsatb\\
  \left[\Nsatb\log(1+\frac{N_b}{\Nsatb}\frac{\alpha_bP_t}{\sgs_b})
 - \Nsate\log\left(1+\frac{\alpha_eP_t}{\beta_eP_j}\right)\right]^+\;,
 \\\text{otherwise.}
  \end{array}
  \right.
\end{equation}
Similar to the case of basic jammer-assisted system, we note that
a non-zero worst-case secrecy capacity sometimes is achievable for
the AN jammer-assisted system, depending on the system parameters, such as transmit power,
average channel gains, the spatial constraint at Bob and the number of antennas at Bob.
Consider the scenario where the number of antennas at Bob, $N_b$, is controllable and the other system parameters
, i.e., $\Nsatb, \Nsate, \alpha_b, \beta_b, \alpha_e, \beta_e, P_t$ and $P_j$, are fixed.
From~\eqref{eq:Cs_WorstM2C2}, we find that a non-zero worst-case secrecy capacity is always achievable by having ``enough" receive antennas at Bob, and the minimum $N_b$ to ensure a non-zero worst-case secrecy capacity is given by
\begin{equation}\label{eq:Nbm_AN}
  \Nbm=\left\{ \begin{array}{ll}
  \left\lfloor\frac{\Nsate\log\left(1+\frac{\alpha_eP_t}{\beta_eP_j}\right)}{\log\left(1+\frac{\alpha_bP_t}{\sgs_b}\right)}\right\rfloor+1\;, \\\text{if}~\Nsatb\log\left(1+\frac{\alpha_bP_t}{\sgs_b}\right)
 \ge\Nsate\log\left(1+\frac{\alpha_eP_t}{\beta_eP_j}\right) \vspace{1mm}\\
 \left\lfloor \frac{\mathrm{\Nsatb}\, {\mathrm{\sigma}_{b}}^2\, }
  { \mathrm{\alpha}_{b}P_{t}}
  \left({\left(1+\frac{\mathrm{\alpha}_{e}P_{t}}{\mathrm{\beta}_{e}P_{j}} \right)}^{\frac{\Nsate}{\mathrm{\Nsatb}}} - 1\right)\right\rfloor+1\;,
  \\\text{otherwise.}
  \end{array}
  \right.
\end{equation}

In terms of the optimal jammer power that maximizes the worst-case secrecy capacity,
it is wise to have $P_j$ as large as possible,
since the increase of $P_j$ only degrades the capacity of Eve's channel but does not affect the capacity of Bob's channel.
Mathematically, we give the following proof for that the worst-case secrecy capacity of the AN jammer-assisted system is a monotonically increasing function of the jamming power.
\begin{IEEEproof}
We first rewrite~\eqref{eq:Cs_WorstM2C2} as
\begin{equation}\label{eq:ANincreaseproof1e}
  C^\wor_{s}
  = \left\{ \begin{array}{ll}
  \left[f_1(P_j)\right]^+\;,
  &\text{if}~N_b\le\Nsatb\\
  \left[f_2(P_j)\right]^+\;, & \text{otherwise.}
  \end{array}
  \right.
\end{equation}
Then, we find that
\begin{equation}\label{eq:ANincreaseproof2e}
  \frac{\partial f_1(P_j)}{\partial P_j}= \frac{\partial f_2(P_j)}{\partial P_j}=\frac{\Nsate P_t \alpha_e}{\left(1+\frac{\alpha_eP_t}{\beta_eP_j}\right)\ln2\beta_eP_j^2}>0
\end{equation}
always holds for any positive value of $P_j$.
Thus, the secrecy capacity of the AN jammer-assisted system is a monotonically increasing function of the jamming power.
\end{IEEEproof}

\subsection{Numerical Results}
In this subsection, we present the numerical results based on the worst-case analysis.
Since the worst-case secrecy capacity of the wiretap-channel system is always equal to zero, we do not present the numerical results for the wiretap-channel system in this subsection but focus on the basic jammer-assisted system and the AN jammer-assisted system.
Besides, we still adopt the 2D analysis to characterize the spatial constraints at the receiver side, such that Bob and Eve are spatially constrained by circular apertures.

\begin{figure}[!t]
\centering
\vspace{-0mm}
\includegraphics[height=2.8in,width=3.3in]{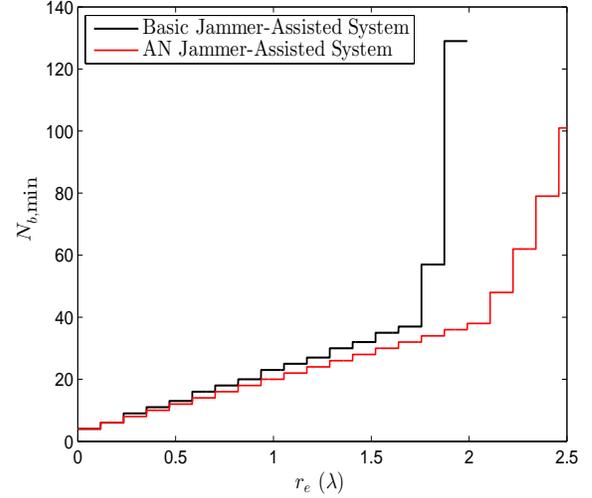}
\vspace{0mm}
\caption{The minimum number of Bob's antennas for achieving a non-zero worst-case secrecy capacity versus the radius of Eve's spatial constraint. The other system parameters are $P_t=20~\mbox{dB}, P_j=0~\mbox{dB}, \alpha_b=1, \alpha_e=1, \beta_b=1, \beta_e=1, \sigma_b^2=1~\mbox{and}~r_b=2\lambda$.}
\vspace{0mm}  \label{fig:NbVsRe}
\end{figure}

We first compare the minimum numbers of Bob's antennas to achieve a non-zero worst-case secrecy capacity of the basic jammer-assisted system and the AN jammer-assisted system.
Figure~\ref{fig:NbVsRe} plots $\Nbm$ versus $r_e$ based on \eqref{eq:Nbm_basic} and \eqref{eq:Nbm_AN}. As shown in the figure, $\Nbm$ increases with $r_e$ for both systems, which indicates that we need more antennas at Bob to ensure a non-zero worst-case secrecy capacity as the radius of Eve's spatial constraint increases.
In addition, we note that the increase of $\Nbm$ with respect to $r_e$ is slow when $r_e$ is small, but it becomes fast when $r_e$ is large. Such an observation is more clear for the basic jammer-assisted system compared with that for the AN jammer-assisted system.
Hence, the cost of antennas at Bob to ensure a non-zero worst-case secrecy capacity is very large when the radius of Eve's spatial constraint is large, especially for the basic jammer-assisted system.
When $r_e$ is very large, i.e., $r_e>r_b=2\lambda$ in the figure, the basic jammer-assisted system cannot achieve a non-zero worst-case secrecy capacity no matter how many antennas are equipped at Bob. The condition under which the basic jammer-assisted system always cannot achieve the non-zero worst-case secrecy capacity is given by \eqref{eq:NbinfinityCs0}.
In contrast, the AN jammer-assisted system can always ensure a non-zero worst-case secrecy capacity by increasing the number of Bob's antennas, as long as Eve has a finite spatial constraint.

It is worth pointing out that the minimum number of receive antennas to ensure a non-zero worst-case secrecy capacity is determined by not only the spatial constraint at the eavesdropper but also many other system parameters, such as the spatial constraint at the legitimate receiver, transmit power, jamming power, average channel gains and the noise variance at the receiver. Thus, the result in Figure~\ref{fig:NbVsRe} can be only regarded as an example to illustrate the required values of $\Nbm$ for different values of $r_e$. The required $\Nbm$ is not necessary to be extremely large for a very large value of $r_e$. For example, the required $\Nbm$ is equal to 116 for $r_e=10\lambda$ in an AN jammer-assisted system with $r_b=8\lambda, \alpha_b=10, \alpha_e=10, \beta_b=10$ and $\beta_e=10$.

\begin{figure}[!t]
\centering
\vspace{-0mm}
\includegraphics[height=2.8in,width=3.3in]{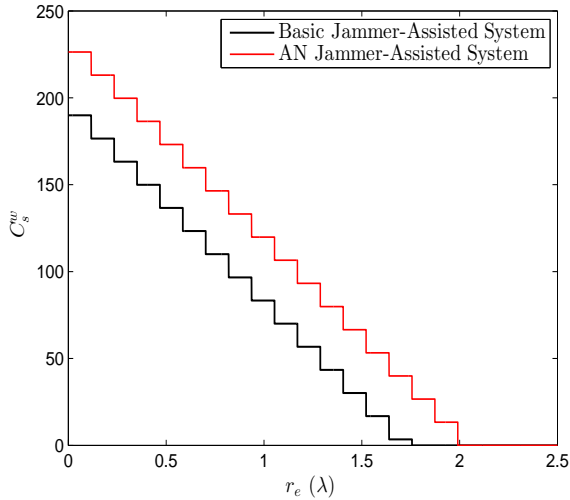}
\vspace{0mm}
\caption{The worst-case secrecy capacity versus the radius of Eve's spatial constraint. The other system parameters are $P_t=20~\mbox{dB}, P_j=0~\mbox{dB}, \alpha_b=1, \alpha_e=1, \beta_b=1, \beta_e=1, \sigma_b^2=1, r_b=2\lambda~\mbox{and}~N_b=\Nsatb=37$.}
\vspace{0mm}  \label{fig:CsVsRe}
\end{figure}

Now, we depict the worst-case secrecy capacity for different spatial constraints at Eve. Figure~\ref{fig:CsVsRe} plots $C_s^w$ versus $r_e$ for the basic jammer-assisted system and the AN jammer-assisted system according to \eqref{eq:Cs_WorstM2C1}  and  \eqref{eq:Cs_WorstM2C2}, respectively. The number of Bob's antennas is chosen equal to the saturation number of receive antennas at Bob, i.e., $N_b=\Nsatb=37$.
As the figure shows, $C_s^w$ decreases with $r_e$ for both systems. 
Comparing the two curves, we note that the worst-case secrecy capacity of the basic jammer-assisted system is always smaller than that for the AN jammer-assisted system. In addition, the difference of $C_s^w$ between the two systems keeps the same for different values of $r_e$.
This can be explained as follows. The basic jamming signals and the AN jamming signals have the same effect on Eve's channel while different effects on Bob's channel. Hence, the difference of $C_s^w$ between the two systems is actually due to the difference of the capacity of Bob's channel subject to different jamming techniques, and it is not related to Eve's channel condition or spatial constraint.
Therefore, the difference of $C_s^w$ between the two curves in the figure keeps the same for different values of $r_e$.

\begin{figure}[!t]
\centering
\vspace{-0mm}
\includegraphics[height=2.8in,width=3.3in]{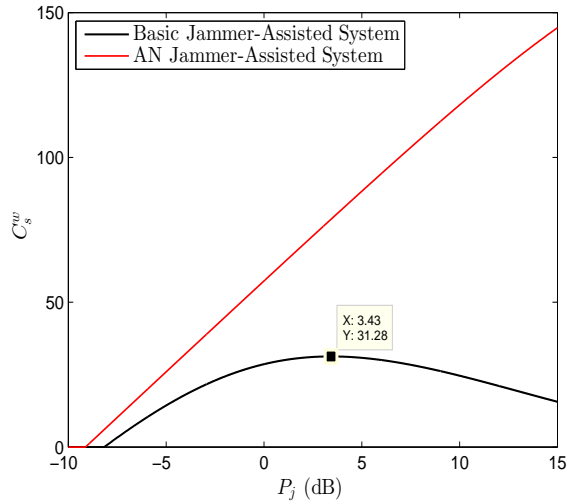}
\vspace{0mm}
\caption{The worst-case secrecy capacity versus the jamming power. The other system parameters are $P_t=20~\mbox{dB}, \alpha_b=1, \alpha_e=1, \beta_b=1, \beta_e=1, \sigma_b^2=1, r_b=1.5\lambda, r_e=1\lambda~\mbox{and}~N_b=30$.}
\vspace{-0mm}  \label{fig:CsVsPj}
\end{figure}

Finally, we illustrate the impact of jamming power on the worst-case secrecy capacity. Figure~\ref{fig:CsVsPj} plots $C_s^w$ versus $P_j$ for both the basic jammer-assisted system and the AN jammer-assisted system. As shown in the figure, the value of $C_s^w$ for the basic jammer-assisted system increases with $P_j$ when $P_j$ is small, but it decreases with $P_j$ when $P_j$ goes large. There exists an optimal value of $P_j$ that maximizes $C_s^w$ for the basic jammer-assisted system, i.e., $P_j=3.43$~dB in the figure. By using the analytical results given in \emph{Proposition~\ref{Pro:Opt_Pj}}, we also obtain that $P_j^\opt$ for the given scenario is equal to $3.43$~dB. This verifies the optimality of $P_j^\opt$ obtained in our analytical results.
In contrast, the value of $C_s^w$ for the AN jammer-assisted system always increases with $P_j$, which is also consistent with our analytical results.
Moreover, comparing the basic jammer-assisted system and the AN jammer-assisted system,
we note that the difference of $C_s^w$ between the two curves increases with $P_j$ all the time.

\section{Conclusion and Future Work}\label{sec:Concl}
In this work, we introduced the spatial constraint into physical layer security for multi-antenna systems, which provides an approach to study the secrecy capacity without knowing the number of eavesdropper antennas.
We considered basic secure communication systems with spatial constraints at the receiver side.
Specifically, we studied the wiretap-channel system, the basic jammer-assisted system and the AN jammer-assisted system, and derived the expressions for secrecy capacity of each system.
We found that a non-zero worst-case secrecy capacity is achievable with the assist of jamming signals, even if the eavesdropper is equipped with infinite number of antennas.
Moreover, the optimal jamming power that maximizes the worst-case secrecy capacity was obtained.
We highlight that the major contribution of this paper is to address the practically important  problem of how to study secure communications without knowing the number of eavesdropper antennas, and hope this work can be a good inspiration for future researchers to design novel physical layer techniques to efficiently secure wireless communications without the information of eavesdropper antennas.

As a first step of studying the effects of spatial constraints on physical layer security, this work considered a simple scenario with spatial constraints at the receiver side only.
A natural future work is to extend the study by investigating the effects of spatial constraints at both the transmitter and the receiver sides.
To this end, a limited number of transmit antennas with the spatial constraint at the transmitter should be considered.
However, it is worth mentioning that the extension is non-trivial, since the secrecy capacity would depend on instantaneous channel realizations even if the number of transmit antennas goes to infinity.
In the scenario with spatial constraints at both the transmitter and the receiver sides, the study of having legitimate CSI available at the transmitter side is also of great interest, while this paper assumed the legitimate CSI available only at the receiver side.
With the legitimate CSI available at the transmitter, how to optimally design the transmit precoding is an interesting problem to investigate.
Another future work direction is to evaluate the achievable secrecy capacity for different antenna array configurations from a signal-processing perspective. Note that the results in this paper were mainly obtained from an information-theoretic perspective by assuming the optimal antenna placement at the receiver side.

\appendices
\section{Proof of Proposition~\ref{Pro:CsM1}}\label{sec:ProofPreli}
\begin{figure}[t!]
\centering
\vspace{-0mm}
\includegraphics[height=2.8in,width=3.3in]{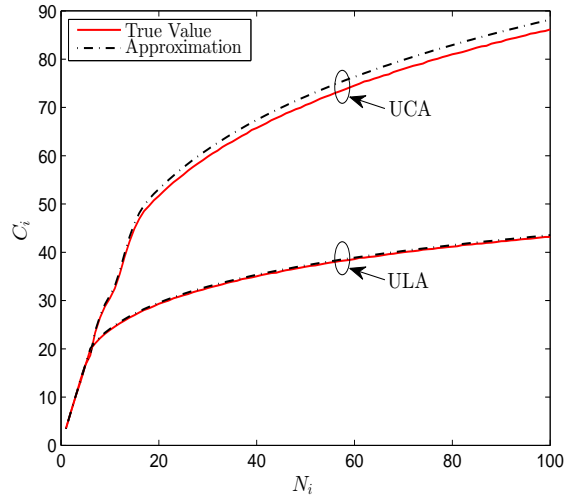}
\vspace{-0mm}
\caption{Without jamming signals: $C_i$ versus $N_i$. The other system parameters are $N_t=100, r_i=1\lambda, P_t=10$ dB, $\alpha_i=1, \sigma_i^2=1$.
}
\vspace{-0mm} \label{fig:AccuracySaNoJamming}
\end{figure}

The capacity of Bob or Eve's channel can be written as
\begin{equation}\label{}
  C_i=\log\left|\bI_{N_i}+\frac{\alpha_i\bH_i\bQ_x\bH_i^H}{\sgs_i}\right|,
\end{equation}
where $\bQ_x$ denotes the covariance matrix of $\bx$, i.e., $\bQ_x=\mathbb{E}\left\{\bx\bx^H\right\}$.
Since Alice has no instantaneous CSI of Bob and there is sufficient space at Alice for independent transmit antenna allocation, the best transmission strategy is to have the transmit signal vector composed of statistically independent equal power components, each with a Gaussian distribution.
Then, the covariance matrix of $\bx$ is equal to
$\bQ_x=\frac{P_t}{N_t}\bI_{N_t}$, and the channel capacity becomes to
\begin{equation}\label{eq:NojammingTrueCi}
  C_i=\log\left|\bI_{N_i}+\frac{\alpha_iP_t}{\sgs_i N_t}\bH_i\bH_i^H\right|,
\end{equation}
where
\begin{equation}\label{eq:App30}
  \bH_i\bH_i^H=\sum_{t=1}^{N_t}\bh_{it}\bh_{it}^H.
\end{equation}
Considering a large number of transmit antennas ($N_t\rightarrow\infty$) and sufficient space for placing transmit antennas (independent $\bh_{it}$), the correlation matrix at the receiver in \eqref{eq:RRh} becomes to
\begin{equation}\label{eq:R_large}
  \bR_i\rightarrow\frac{1}{N_t}\sum_{t=1}^{N_t}\bh_{it}\bh_{it}^H.
\end{equation}
Note that there is no expectation over channel realizations in \eqref{eq:R_large}, since $\frac{1}{N_t}\sum_{t=1}^{N_t}\bh_{it}\bh_{it}^H=\mathbb{E}\left\{\frac{1}{N_t}\sum_{t=1}^{N_t}\bh_{it}\bh_{it}^H\right\}$ when $N_t\rightarrow\infty$.
Then, the channel capacity with a large number of sufficiently separated transmit antennas is approximated by
\begin{equation}\label{eq:NojammingAppSa}
  C_i\approx\log\left|\bI_{N_i}+\frac{\alpha_iP_t}{\sgs_i}\bR_i\right|.
\end{equation}
We highlight that the approximation by \eqref{eq:NojammingAppSa} provides good accuracy even if the number of transmit antennas is finite.
To examine the accuracy of the approximation by~\eqref{eq:NojammingAppSa}, we compare the true value of $C_i$ obtained by  \eqref{eq:NojammingTrueCi} and the approximation obtain by \eqref{eq:NojammingAppSa} for given receive antenna array configurations.
The simulation result is presented by Figure~\ref{fig:AccuracySaNoJamming}. The number of transmit antennas is set as a large but finite number, $N_t=100$. The number of receive antennas is in the range of $1\le N_i\le N_t=100$.
We consider two different antenna array configurations, which are the uniform linear array (ULA) and the uniform circular array (UCA), in a fixed circular aperture at the receiver with $r_i=1\lambda$.
Since the number of transmit antennas is set as a finitely large number but not infinity, the capacity result by \eqref{eq:NojammingTrueCi} would depend on the instantaneous channel realization. Thus, the ``true value" in Figure~\ref{fig:AccuracySaNoJamming} is the average value of $C_i$ obtained by \eqref{eq:NojammingTrueCi} over different channel realizations.
It is evident from Figure~\ref{fig:AccuracySaNoJamming} that the difference between the true value and the approximation is very small for the whole range of $N_i$, which indicates that the approximation by \eqref{eq:NojammingAppSa} provides good accuracy even if the transmitter has a finite number of antennas.

For the receiver with $N_i$ optimally-placed antennas in a fixed aperture region, the channel capacity in~\eqref{eq:NojammingAppSa} can be further approximated by~\cite[Chapter~3]{Pollock_03_On},
\begin{equation}\label{eq:Ci}
  C_i \approx \left\{ \begin{array}{ll} N_i\log(1+\frac{\alpha_iP_t}{\sgs_i})\;, &\text{if}~N_i\le\Nsati \\
              \Nsati\log(1+\frac{N_i}{\Nsati}\frac{\alpha_iP_t}{\sgs_i})\;, &\text{otherwise,}
  \end{array}
  \right.
\end{equation}
where the expression of $\Nsati$ for a 2D circular aperture or
a 3D spherical aperture is given by~\eqref{eq:Nsati}.
The $C_i$ in~\eqref{eq:Ci} is derived with the approximation that
$ J_m\left(\frac{2\pi}{\lambda}r_i\right)\rightarrow 0$
for $m\ge\left\lceil\pi e r_i/\lam \right\rceil+1$, where $J_m(\cdot)$ denotes the Bessel function of order $m$. Such an approximation is shown to be very accurate in~\cite{Pollock_03_On}.

Finally, substituting~\eqref{eq:Ci} into~\eqref{eq:CsB} completes the proof of Proposition~\ref{Pro:CsM1}.

\section{Proof of Theorem~\ref{Theo:CsM2C1}}\label{sec:ProofCsM2C1}
The capacity of Bob or Eve's channel subject to the basic jamming signals is written as~\cite[Section 3.1]{Bayesteh_04_Effect}
\begin{equation}\label{eq:Bayesteh_04_Effect}
  C_i=\log\left|\bI_{N_i}+\alpha_i\bH_i\bQ_x\bH_i^H\left(\beta_i\bG_i\bQ_w\bG_i^H+\sgs_i\bI_{N_i}\right)^{-1}\right|,
\end{equation}
where $\bQ_x$ and $\bQ_w$ denote the covariance matrices of $\bx$ and $\bw_1$, respectively, i.e., $\bQ_x=\mathbb{E}\left\{\bx\bx^H\right\}$ and $\bQ_w=\mathbb{E}\left\{\bw_1\bw_1^H\right\}$. Since neither Alice nor Helen has the instantaneous CSI to Bob or Eve, the equal power allocation at the transmit antennas is adopted at both Alice and Helen, and the covariance matrices of $\bx$ and $\bw_1$ are equal to
$\bQ_x=\frac{P_t}{N_t}\bI_{N_t}$ and $\bQ_w=\frac{P_j}{N_j}\bI_{N_j}$, respectively. Then, the channel capacity becomes to
\begin{equation}\label{eq:jammingTrueCi}
  C_i=\log\left|\bI_{N_i}+\frac{\alpha_iP_t}{N_t}\bH_i\bH_i^H\left(\frac{\beta_iP_j}{N_j}\bG_i\bG_i^H+\sgs_i\bI_{N_i}\right)^{-1}\right|.
\end{equation}
Considering the large number of transmit antennas ($N_t\rightarrow\infty, N_j\rightarrow\infty$) and sufficient space for placing transmit antennas (independent $\bh_{it}$ and independent $\bg_{it}$), we have
\begin{equation}\label{eq:what41}
 \frac{1}{N_t}\sum_{t=1}^{N_t}\bh_{it}\bh_{it}^H=\frac{1}{N_j}\sum_{t=1}^{N_j}\bg_{it}\bg_{it}^H=\bR_i,
\end{equation}
where $\bR_i$ is the correlation matrix at the receiver side. Note that $\bR_i$ is determined by the receive antenna correlations.

\begin{figure}[t!]
\centering
\vspace{-0mm}
\includegraphics[height=2.8in,width=3.3in]{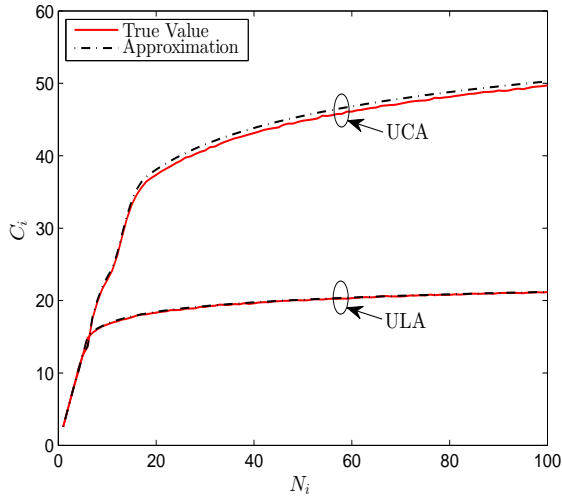}
\vspace{-0mm} \caption{With jamming signals: $C_i$ versus $N_i$. The other system parameters are $N_t=N_j=100, r_i=1\lambda, P_t=10$ dB, $P_j=0$ dB $\alpha_i=1, \beta_i=1, \sigma_i^2=1$.} \label{fig:AccuracySaJamming}
\vspace{-0mm}
\end{figure}

Therefore, the channel capacity can be approximated by
\begin{eqnarray}\label{eq:jammingAppSa}
  C_i &\approx& \log\left|\bI_{N_i}+\alpha_iP_t\bR_i\left(\beta_iP_j\bR_i+\sgs_i\bI_{N_i}\right)^{-1}\right|\nonumber\\
   &=& \log\left|\bI_{N_i}+\left(\frac{\alpha_iP_t}{\sgs_i}+\frac{\beta_iP_j}{\sgs_i}\right)\bR_i\right|\nonumber\\
   && -\log\left|\bI_{N_i}+\frac{\beta_iP_j}{\sgs_i}\bR_i\right|.
\end{eqnarray}
We highlight that the approximation by \eqref{eq:jammingAppSa} provides good accuracy even if the number of transmit antennas and the number of jamming antennas are finite.
To examine the accuracy of the approximation by~\eqref{eq:jammingAppSa}, we compare the true value of $C_i$ obtained by  \eqref{eq:jammingTrueCi} and the approximation obtain by \eqref{eq:jammingAppSa} for given receive antenna array configurations. The simulation result is presented by Figure~\ref{fig:AccuracySaJamming}.
The number of transmit antennas and the number of jamming antennas are set as $N_t=N_j=100$. The number of receive antennas is in the range of $1\le N_i\le N_t=N_j=100$. We still consider two different antenna array configurations, i.e., the ULA and the UCA, in a fixed circular aperture at the receiver with $r_i=1\lambda$.
It is evident from Figure~\ref{fig:AccuracySaJamming} that the difference between the true value and the approximation is very small for the whole range of $N_i$. This confirms that the approximation by \eqref{eq:jammingAppSa} provides good accuracy even if the transmitter and the jammer have finite numbers of antennas.

For the receiver with $N_i$ optimally-placed antennas in a fixed aperture region, the channel capacity in~\eqref{eq:jammingAppSa} can be further approximated by
\begin{equation}\label{eq:CiM2}
    C_i \approx
  \left\{\begin{array}{ll} N_i\log\left(1+\frac{\alpha_iP_t}{\beta_iP_j+\sgs_i}\right)\;, &\text{if}~N_i\le\Nsati\\
  \Nsati\log\left(1+\frac{\frac{N_i}{\Nsati}\alpha_iP_t}{\frac{N_i}{\Nsati}\beta_iP_j+\sgs_i}\right)\;,
  &\text{otherwise.}\end{array}
  \right.
\end{equation}
Still, the $C_i$ in~\eqref{eq:CiM2} is derived with the approximation that
$J_m\left(\frac{2\pi}{\lambda}r_i\right)\rightarrow 0$ for $m\ge\left\lceil\pi e r_i/\lam \right\rceil+1$.

Finally, substituting~\eqref{eq:CiM2} into~\eqref{eq:CsB} completes the proof of Theorem~\ref{Theo:CsM2C1}.

\section{Proof of Theorem~\ref{Theo:CsM2C2}}\label{sec:ProofCsM2C2}
Since the AN jamming signals do not degrade Bob's channel, we derive the capacity of Bob's channel directly from~\eqref{eq:Ci}, which is given by
\begin{equation}\label{eq:CbM2C2}
  C_b \approx \left\{ \begin{array}{ll} N_b\log(1+\frac{\alpha_bP_t}{\sgs_b})\;, &\text{if}~N_b\le\Nsatb\\
  \Nsatb\log(1+\frac{N_b}{\Nsatb}\frac{\alpha_bP_t}{\sgs_b})\;, & \text{otherwise.}
  \end{array}
  \right.
\end{equation}

Now, we derive the capacity of Eve's channel subject to the AN jamming signals.
The received signal vector at Eve is written as
\begin{equation}\label{}
  \by_e=\sqrt{\alpha_e}\bH_e\bx+\sqrt{\beta_e}\bK\bv+\bn_e,
\end{equation}
where $\bK=\bG_e\bZ$ represents the equivalent channel for the vector $\bv$ to Eve. Due to the orthonormality of $\bZ$, the $N_e\times(N_j-N_b)$ matrix $\bK$ has circularly symmetric i.i.d. complex Gaussian distributed elements.
Then, the capacity of Eve's channel is written as
\begin{equation}\label{}
  C_e=\log\left|\bI_{N_e}+\alpha_e\bH_e\bQ_x\bH_e^H\left(\beta_e\bK\bQ_v\bK^H+\sgs_e\bI_{N_e}\right)^{-1}\right|,
\end{equation}
where $\bQ_x$ and $\bQ_v$ denote the covariance matrices of $\bx$ and $\bv$, respectively, i.e., $\bQ_x=\mathbb{E}\left\{\bx\bx^H\right\}$ and $\bQ_v=\mathbb{E}\left\{\bv\bv^H\right\}$.
With the equal power allocation at Alice, we have
$\bQ_x=\frac{P_t}{N_t}\bI_{N_t}$. Also, since $\bv$ is chosen as i.i.d. complex Gaussian random variables, we have $\bQ_v=\frac{P_j}{N_j-N_b}\bI_{N_j-N_b}$.
Then, the capacity of Eve's channel becomes to
\begin{equation}\label{}
  C_e=\log\left|\bI_{N_e}+\alpha_eP_t\bR_e\left(\frac{\beta_eP_j}{N_j-N_b}\bK\bK^H+\sgs_e\bI_{N_e}\right)^{-1}\right|,
\end{equation}
where $\bR_e$ is the correlation matrix at Eve, and is determined by the receive antenna correlations at Eve.
Define $\bK=[\bk_{1} \cdots \bk_i \cdots \bk_{N_j-N_b}]$,
$\bZ=[\bz_{1} \cdots \bz_i \cdots \bz_{N_j-N_b}]$, and hence $\bk_i=\bH_e\bz_i$.

\emph{If we can prove that $\bk_i$ are independent}, the correlation matrix would converge to $\bR\rightarrow\frac{1}{N_j-N_b}\bK\bK^H$ as $(N_j-N_b)\rightarrow\infty$, and the capacity of Eve's channel could be written as
\begin{equation}\label{eq:CeM2C2s1}
  C_e=\log\left|\bI_{N_e}+\alpha_eP_t\bR_e\left(\beta_eP_j\bR_e+\sgs_e\bI_{N_e}\right)^{-1}\right|.
\end{equation}
Having~\eqref{eq:CeM2C2s1}, we can derive the channel capacity of spatially-constrained Eve which is the same as~\eqref{eq:CiM2}.

Therefore, in the following, we need only to prove that $\bk_i$ are independent to complete the proof of Theorem~\ref{Theo:CsM2C2}.
For any $\bk_m$ and $\bk_n$ where $m\neq n$, we have
\begin{equation}\label{}
  [\bk_m-\mathbb{E}\left\{\bk_m\right\}]^H[\bk_n-\mathbb{E}\left\{\bk_n\right\}] = \bz_m^H\bH_e^H\bH_e\bz_n\stackrel{\text{(a)}}{=}\bz_m^H\bz_n \stackrel{\text{(b)}}{=}0,
\end{equation}
where (a) is because of the independence between transmit antennas and (b) is because of the orthogonality of $\bZ$.
Thus, $\bk_i$ are pairwise uncorrelated.
In addition, multivariate normality and no correlation implies independence. Multivariate normality
and pairwise independence implies mutual independence. Since $\bk_i$ are multivariate normally distributed, $\bk_i$ are mutually independent.
This completes the proof of Theorem~\ref{Theo:CsM2C2}.

\section{Proof of Proposition~\ref{Pro:Opt_Pj}}\label{sec:ProofOptPj}
We first rewrite~\eqref{eq:Cs_WorstM2C1} as
\begin{equation}\label{}
  C^\wor_{s}
  = \left\{ \begin{array}{ll}
  \left[f_1(x=P_j)\right]^+\;,
  &\text{if}~N_b\le\Nsatb\\
  \left[f_2(x=P_j)\right]^+\;, & \text{otherwise.}
  \end{array}
  \right.
\end{equation}
If $N_b\le\Nsatb$, we can obtain two possible stationary points of $f_1(x)$, i.e., $x_1$ and $x_2$, by taking the derivative of $f_1(x)$ with respect to $x$ and equating it to zero.
If $C^\wor_{s}$ is not always equal to zero, $P_j^\opt$ should exist and be equal to one of the stationary points,
since $\lim_{x\rightarrow0}f(x)\rightarrow-\infty$ and $\lim_{x\rightarrow\infty}f(x)\rightarrow0$.
Then, we determine $P_j^\opt$ by examining the values of $x_1$ and $x_2$.
When neither $x_1$ nor $x_2$ is real and positive, it is not applicable to determine the optimal value of $P_j$, because the stationary point for $f_1(x)$ does not exist, and $C^\wor_{s}$ is always equal to zero for any value of $P_j$.
Similarly, if $N_b>\Nsatb$, we can obtain two possible stationary points of $f_2(x)$, i.e., $x_3$ and $x_4$, by taking the derivative of $f_2(x)$ with respect to $x$ and equating it to zero.
Then, we determine $P_j^\opt$ by examining the values of $x_3$ and $x_4$. When neither $x_3$ nor $x_4$ is real and positive, it is not applicable to determine the optimal value of $P_j$, because  $C^\wor_{s}$ is always equal to zero for any value of $P_j$.
This completes the proof of Proposition~\ref{Pro:Opt_Pj}.


\balance


\begin{thebibliography}{10}
\providecommand{\url}[1]{#1}
\csname url@samestyle\endcsname
\providecommand{\newblock}{\relax}
\providecommand{\bibinfo}[2]{#2}
\providecommand{\BIBentrySTDinterwordspacing}{\spaceskip=0pt\relax}
\providecommand{\BIBentryALTinterwordstretchfactor}{4}
\providecommand{\BIBentryALTinterwordspacing}{\spaceskip=\fontdimen2\font plus
\BIBentryALTinterwordstretchfactor\fontdimen3\font minus
  \fontdimen4\font\relax}
\providecommand{\BIBforeignlanguage}[2]{{%
\expandafter\ifx\csname l@#1\endcsname\relax
\typeout{** WARNING: IEEEtran.bst: No hyphenation pattern has been}%
\typeout{** loaded for the language `#1'. Using the pattern for}%
\typeout{** the default language instead.}%
\else
\language=\csname l@#1\endcsname
\fi
#2}}
\providecommand{\BIBdecl}{\relax}
\BIBdecl

\bibitem{Bloch_11}
M.~Bloch and J.~Barros, \emph{Physical-Layer Security: From Information Theory
  to Security Engineering}.\hskip 1em plus 0.5em minus 0.4em\relax Cambridge
  University Press, 2011.

\bibitem{Zhou_13_Physical}
X.~Zhou, L.~Song, and Y.~Zhang, \emph{Physical Layer Security in Wireless
  Communications}.\hskip 1em plus 0.5em minus 0.4em\relax CRC Press, 2013.

\bibitem{wyner_75}
A.~D. Wyner, ``The wire-tap channel,'' \emph{Bell Syst. Tech. J.}, vol.~54,
  no.~8, pp. 1355--1387, Oct. 1975.

\bibitem{csiszar_78}
I.~Csisz\'{a}r and J.~K\"{o}rner, ``Broadcast channels with confidential
  messages,'' \emph{IEEE Trans. Inf. Theory}, vol.~24, no.~3, pp. 339--348, May
  1978.

\bibitem{Cheong_78}
S.~K. Leung-Yan-Cheong and M.~E. Hellman, ``The {G}aussian wire-tap channel,''
  \emph{IEEE Trans. Inf. Theory}, vol.~24, no.~4, pp. 451--456, July 1978.

\bibitem{Khisti_10}
A.~Khisti and G.~W. Wornell, ``Secure transmission with multiple antennas {I}:
  The {MISOME} wiretap channel,'' \emph{IEEE Trans. Inf. Theory}, vol.~56,
  no.~7, pp. 3088--3104, July 2010.

\bibitem{Khisti_10_II}
------, ``Secure transmission with multiple antennas--{P}art {II}: The {MIMOME}
  wiretap channel,'' \emph{IEEE Trans. Inf. Theory}, vol.~56, no.~11, pp.
  5515--5532, Nov. 2010.

\bibitem{Oggier_11_The}
F.~Oggier and B.~Hassibi, ``The secrecy capacity of the {MIMO} wiretap
  channel,'' \emph{IEEE Trans. Inf. Theory}, vol.~57, no.~8, pp. 4961--4972,
  Aug. 2011.

\bibitem{Goel_08}
S.~Goel and R.~Negi, ``Guaranteeing secrecy using artificial noise,''
  \emph{IEEE Trans. Wireless Commun.}, vol.~7, no.~6, pp. 2180--2189, June
  2008.

\bibitem{Zhou_10}
X.~Zhou and M.~R. McKay, ``Secure transmission with artificial noise over
  fading channels: Achievable rate and optimal power allocation,'' \emph{IEEE
  Trans. Veh. Technol.}, vol.~59, no.~8, pp. 3831--3842, Oct. 2010.

\bibitem{Zhou_11_secure}
X.~Zhou, R.~K. Ganti, and J.~G. Andrews, ``Secure wireless network connectivity
  with multi-antenna transmission,'' \emph{IEEE Trans. Wireless Commun.},
  vol.~10, no.~2, pp. 425--430, Feb. 2011.

\bibitem{Mukherjee_11}
A.~Mukherjee and A.~L. Swindlehurst, ``Robust beamforming for security in
  {MIMO} wiretap channels with imperfect {CSI},'' \emph{IEEE Trans. Signal
  Process.}, vol.~59, no.~1, pp. 351--361, Jan. 2011.

\bibitem{Yang_13_Transmit}
N.~Yang, P.~L. Yeoh, M.~Elkashlan, R.~Schober, and I.~B. Collings, ``Transmit
  antenna selection for security enhancement in {MIMO} wiretap channels,''
  \emph{IEEE Trans. Commun.}, vol.~61, no.~1, pp. 144--154, Jan. 2013.

\bibitem{He_13_2}
B.~He and X.~Zhou, ``Secure on-off transmission design with channel estimation
  errors,'' \emph{IEEE Trans. Inf. Forensics Security}, vol.~8, no.~12, pp.
  1923--1936, Dec. 2013.

\bibitem{He_13_3}
B.~He, X.~Zhou, and T.~D. Abhayapala, ``Wireless physical layer security with
  imperfect channel state information: A survey,'' \emph{ZTE Commun.}, vol.~11,
  no.~3, pp. 11--19, Sept. 2013.

\bibitem{Cumanan_14_Secrecy}
K.~Cumanan, Z.~Ding, B.~Sharif, G.~Y. Tian, and K.~K. Leung, ``Secrecy rate
  optimizations for a {MIMO} secrecy channel with a multiple-antenna
  eavesdropper,'' \emph{IEEE Trans. Veh. Technol.}, vol.~63, no.~4, pp.
  1678--1690, May 2014.

\bibitem{Chu_15_Secrecy}
Z.~Chu, K.~Cumanan, Z.~Ding, M.~Johnston, and S.~L. Goff, ``Secrecy rate
  optimizations for a {MIMO} secrecy channel with a cooperative jammer,''
  \emph{IEEE Trans. Veh. Technol.}, vol.~64, no.~5, pp. 1833--1847, May 2015.

\bibitem{Chu_15_Robust}
------, ``Robust outage secrecy rate optimizations for a {MIMO} secrecy
  channel,,'' \emph{IEEE Wireless Commun. Lett.}, vol.~4, no.~1, pp. 86--89,
  Feb. 2015.

\bibitem{Pollock_03_On}
\BIBentryALTinterwordspacing
T.~S. Pollock, ``On limits of multi-antenna wireless communications in
  spatially selective channels,'' Ph.D. dissertation, The Australian National
  University, Australia, July 2003. [Online]. Available:
  \url{http://hdl.handle.net/1885/47999}
\BIBentrySTDinterwordspacing

\bibitem{Larsson_14_Massive}
E.~G. Larsson, O.~Edfors, F.~Tufvesson, and T.~Marzetta, ``Massive {MIMO} for
  next generation wireless systems,'' \emph{IEEE Commun. Mag.}, vol.~52, no.~2,
  pp. 186--195, Feb. 2014.

\bibitem{Andrews_14_What}
J.~G. Andrews, S.~Buzzi, W.~Choi, S.~V. Hanly, A.~Lozano, A.~C.~K. Soong, and
  J.~C. Zhang, ``What will 5{G} be?'' \emph{IEEE J. Sel. Areas Commun.},
  vol.~32, no.~6, pp. 1065--1082, June 2014.

\bibitem{Pollock_03_J_Introducing}
T.~S. Pollock, T.~D. Abhayapala, and R.~A. Kennedy, ``Introducing space into
  {MIMO} capacity calculations,'' \emph{J. Telecommun. Syst.}, vol.~24, no.~2,
  pp. 415--436, Oct. 2003.

\bibitem{Abhayapala_02_On}
T.~D. Abhayapala, R.~A. Kennedy, and J.~T.~Y. Ho, ``On capacity of
  multi-antenna wireless channels: Effects of antenna separation and spatial
  correlation,'' in \emph{Proc. IEEE AusCTW}, Canberra, Australia, Feb. 2002,
  pp. 4--5.

\bibitem{Pollock_03_bound}
T.~S. Pollock, T.~D. Abhayapala, and R.~A. Kennedy, ``Introducing `space' into
  space-time {MIMO} capacity calculations: A new closed form upper bound,'' in
  \emph{Proc. ICT}, vol.~2, Feb. 2003, pp. 1536--1541.

\bibitem{Kennedy_07_Instinsic}
R.~A. Kennedy, P.~Sadeghi, T.~D. Abhayapala, and H.~M. Jones, ``Intrinsic
  limits of dimensionality and richness in random multipath fields,''
  \emph{IEEE Trans. Signal Process.}, vol.~55, no.~6, pp. 2542--2556, June
  2007.

\bibitem{Bashar_12_Degrees}
F.~Bashar and T.~D. Abhayapala, ``Degrees of freedom of band limited signals
  measured over space,'' in \emph{Proc. ISCIT}, Oct. 2012, pp. 735--740.

\bibitem{Bashar_14_Analysis}
F.~Bashar, S.~M.~A. Salehin, and T.~D. Abhayapala, ``Analysis of degrees of
  freedom of wideband random multipath fields observed over time and space
  windows,'' in \emph{IEEE Workshop SSP}, June 2014, pp. 45--48.

\bibitem{Bashar_14_Band}
\BIBentryALTinterwordspacing
------, ``Band limited signals observed over finite spatial and temporal
  windows: An upper bound to signal degrees of freedom,'' submitted to
  \textit{IEEE Trans. Signal Process.}, 2014. [Online]. Available:
  \url{http://arxiv.org/abs/1405.2163}
\BIBentrySTDinterwordspacing

\bibitem{Gesbert_02_Capacity}
D.~Gesbert, T.~Ekman, and N.~Christophersen, ``Capacity limits of dense
  palm-sized {MIMO} arrays,'' in \emph{Proc. IEEE GLOBECOM}, vol.~2, Nov. 2002,
  pp. 1187--1191.

\bibitem{Hanlen_02_Capacity}
L.~Hanlen and M.~Fu, ``Capacity of {MIMO} wireless systems with spatially
  correlated receive elements,'' in \emph{Proc. WITSP}, Wollongong, Australia,
  Dec. 2002, pp. 1--6.

\bibitem{Pollock_03_Antenna}
T.~S. Pollock, T.~D. Abhayapala, and R.~A. Kennedy, ``Antenna saturation
  effects on {MIMO} capacity,'' in \emph{Proc. IEEE ICC}, vol.~4, May 2003, pp.
  2301--2305.

\bibitem{Wu_07_On}
Y.~Wu and Z.~Nie, ``On the {MIMO} channel capacity saturation for spatially
  constrained receive region,'' \emph{J. Systems Engineering Electronics},
  vol.~18, no.~3, pp. 437--442, Sept. 2007.

\bibitem{Bayesteh_04_Effect}
A.~Bayesteh, M.~Ansari, and A.~K. Khandani, ``Effect of jamming on the capacity
  of {MIMO} channels,'' in \emph{Proc. Allerton}, Oct. 2004, pp. 401--410.

\end{thebibliography}
\end{document}